\documentclass[a4paper]{article}
\usepackage{graphicx}
\usepackage{subfigure}
\usepackage{hyperref}

\graphicspath{{./images/}}

\title{The COMPASS Hadron Spectroscopy Programme}

\author{Alexander Austregesilo$^{1,2}$ for the COMPASS Collaboration}

\date{}

\begin{document}

\maketitle

$^1$ Technische Universit\"at M\"unchen, Physik-Department E18, James-Franck-Stra\ss e, 85748~Garching, Germany

$^2$ CERN PH-SME, 1211 Geneva 23, Switzerland

\vspace{12pt}

E-mail address: \href{mailto:Alexander.Austregesilo@cern.ch}{Alexander.Austregesilo@cern.ch}

\vspace{12pt}

\begin{abstract}
COMPASS is a fixed-target experiment at the CERN SPS for the investigation of the structure and the dynamics of hadrons. The experimental setup features a large acceptance and high momentum resolution spectrometer including particle identification and calorimetry and is therefore ideal to access a broad range of different final states.
Following the promising observation of a spin-exotic resonance during an earlier pilot run, COMPASS focused on light-quark hadron spectroscopy during the years 2008 and 2009. A data set, world leading in terms of statistics and resolution, has been collected with a 190\,GeV/$c$ hadron beam impinging on either liquid hydrogen or nuclear targets. Spin-exotic meson and glueball candidates formed in both diffractive dissociation and central production are presently studied. Since the beam composition includes protons, the excited baryon spectrum is also accessible. Furthermore, Primakoff reactions have the potential to determine radiative widths of the resonances and to probe chiral perturbation theory.
An overview of the ongoing analyses will be presented. In particular, the employed partial wave analysis techniques will be illustrated and recent results will be shown for a selection of final states.
\end{abstract}


\section{Introduction}
\subsection{Meson Spectroscopy}

The Constituent Quark Model (CQM) describes hadrons as compositions of quarks and anti-quarks. Due to the self-coupling of colour-charged gluons, additional exotic states are predicted by Quantum Chromo-Dynamics (QCD). Pure gluonic excitations without valence quark content, so-called glueballs, are allowed as well as hybrid mesons. Lattice-QCD predictions \cite{lqcd} expect the lightest glueball to be a scalar particle ($J^{PC}=0^{++}$) with a mass of approximately $1.7\,\mathrm{GeV}/c^2$ and candidates have indeed been observed by a few experiments \cite{cb}\cite{wa102}. However, mixing with isoscalar mesons impedes a clear interpretation. On the other hand, predicted light hybrids with exotic quantum numbers which have a vanishing $q\bar q$ term are promising candidates for physics beyond the CQM, the lightest ones having $J^{PC}=1^{-+}$ and a mass between $1.3$ and $2.2\,\mathrm{GeV}/c^2$ \cite{hyb}.

Two experimentally observed $1^{-+}$ resonances in the light-quark sector have been reported in the past. $\pi_1$(1400) has been observed with $\eta\pi$ decays in \cite{e582}\cite{ves}\cite{cb2}, while the decay of $\pi_1$(1600) into $\rho\pi$\cite{rhopi}\cite{rhopi2}, $\eta'\pi$\cite{ves}\cite{etapi}, $f_1\pi$\cite{f1pi}\cite{f1pi2} and $\omega\pi\pi$\cite{f1pi2}\cite{omegapipi} was each reported by two experiments. In particular, the resonant nature of the $\pi_1$(1600) observed in the three-pion final state is highly disputed\cite{f1pi2}\cite{hyb2}.

Several different production mechanisms can be used in order to produce these resonances experimentally. Diffractive dissociation of the beam particle on an inert target is ideal to study spin-exotic mesons. The $t$-channel Reggeon exchange reaction is characterised by forward kinematics and requires separation of the decay products at very small angles. Central production with double-pomeron exchange on the other hand is especially suited for the search of glue-rich states. Due to the large rapidity gap between the beam particle and the resonance, relatively large angles can occur in this case. Furthermore, the photo-production of mesons can provide a measurement of the radiative widths of the resonances and tests of Chiral Perturbation Theory\cite{chipt}.

\begin{figure}[h]
  \subfigure[]{\includegraphics[width=.32\textwidth]{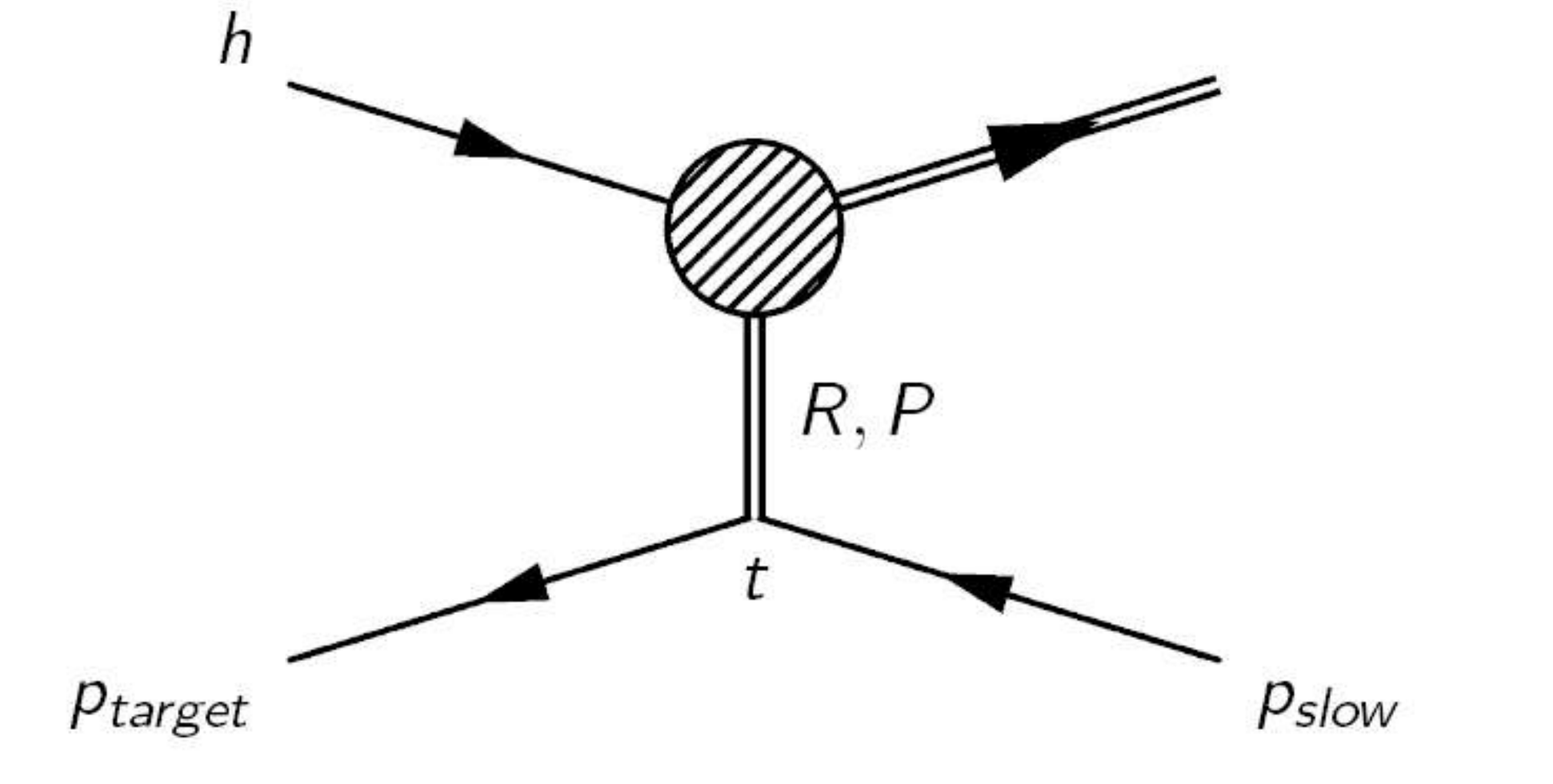}}
  \subfigure[]{\includegraphics[width=.32\textwidth]{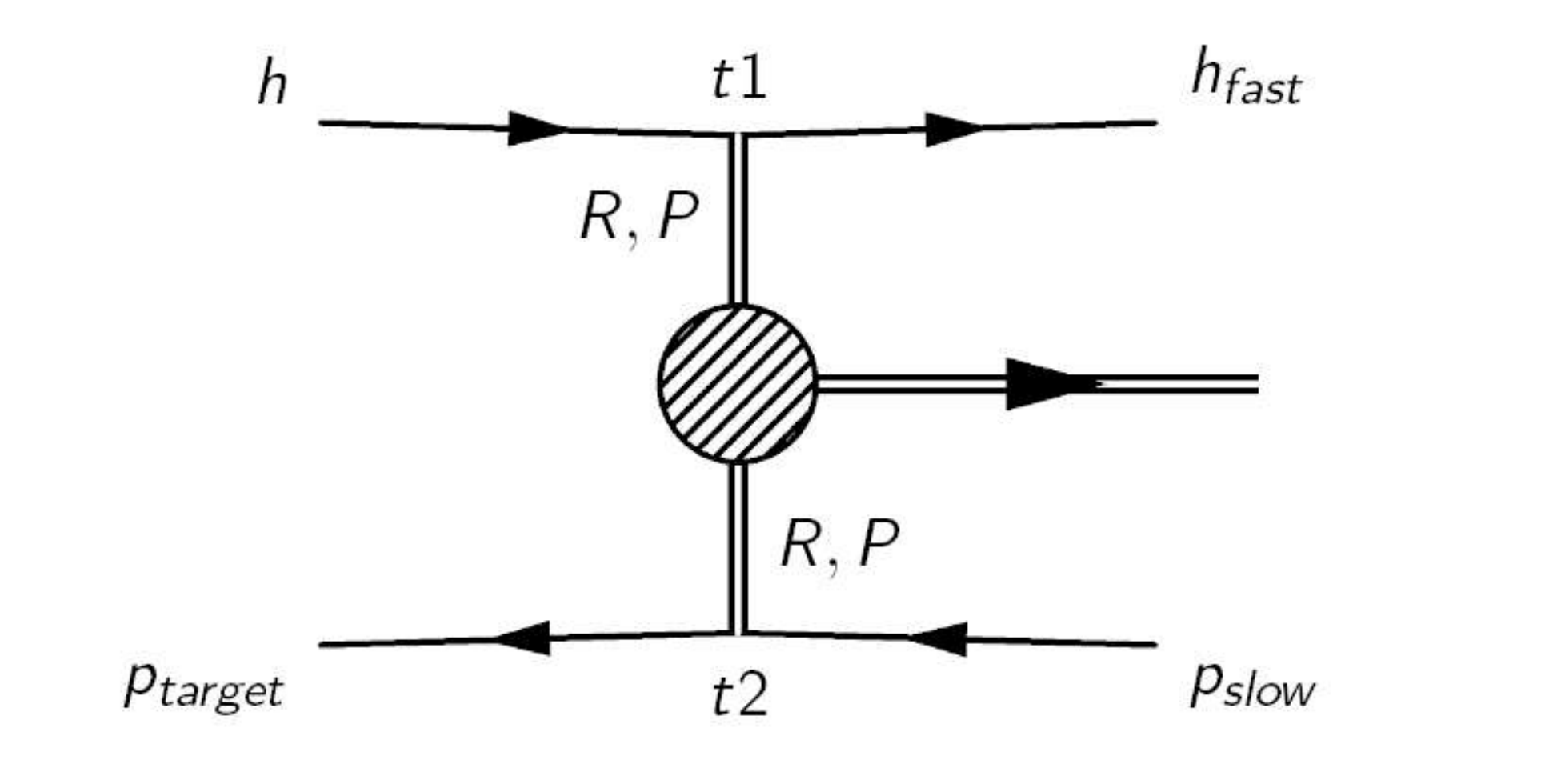}}
  \subfigure[]{\includegraphics[width=.32\textwidth]{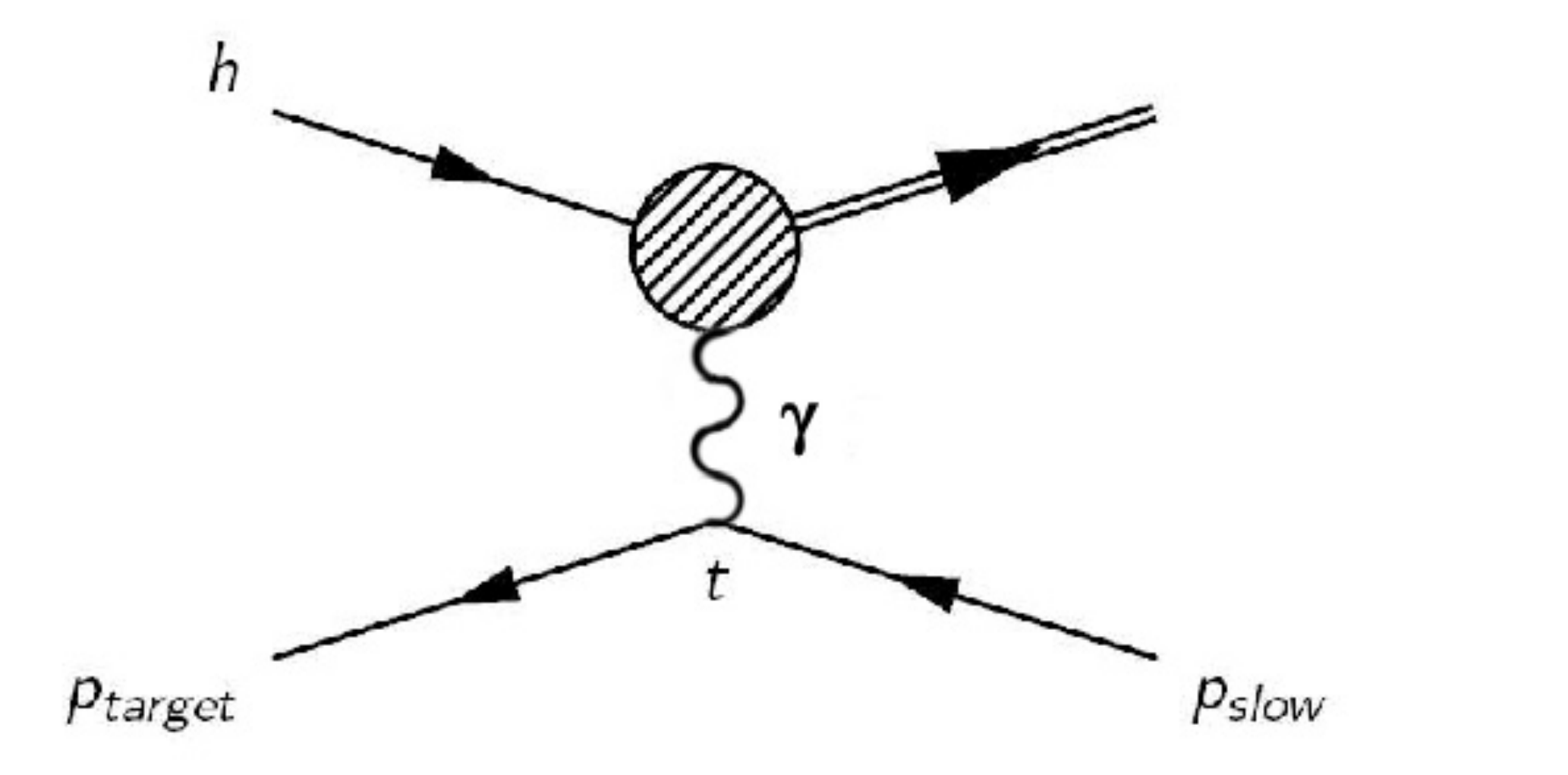}}
  \caption{(a)~{\em Diffractive scattering} (b)~{\em Central production} (c)~{\em Photo-production}}
  \label{fig:tprime2}
\end{figure}

All in all, a competitive experiment has to meet the requirements for precise detection of various decay modes, both charged and neutral, to determine the nature of the resonances. The COMPASS experiment is therefore ideal to explore the field of hadron physics.

\subsection{COMPASS experiment}

The COmmon Muon and Proton Apparatus for Structure and Spectroscopy (COMPASS)\cite{com07} is a fixed target experiment at the CERN SUPER Proton Synchrotron (SPS). It is a two-stage spectrometer that covers a wide range of scattering angles and charged particle momenta with a tracking system which provides high angular resolution. It is equipped with electromagnetic and hadronic calorimeters in both stages to reconstruct also neutral particles in the final state. A Ring Imaging Cherenkov detector (RICH) provides additional particle identification for protons, kaons and pions in a limited kinematic range.

\begin{figure}[h]
  \centering
  \includegraphics[width=.95\textwidth]{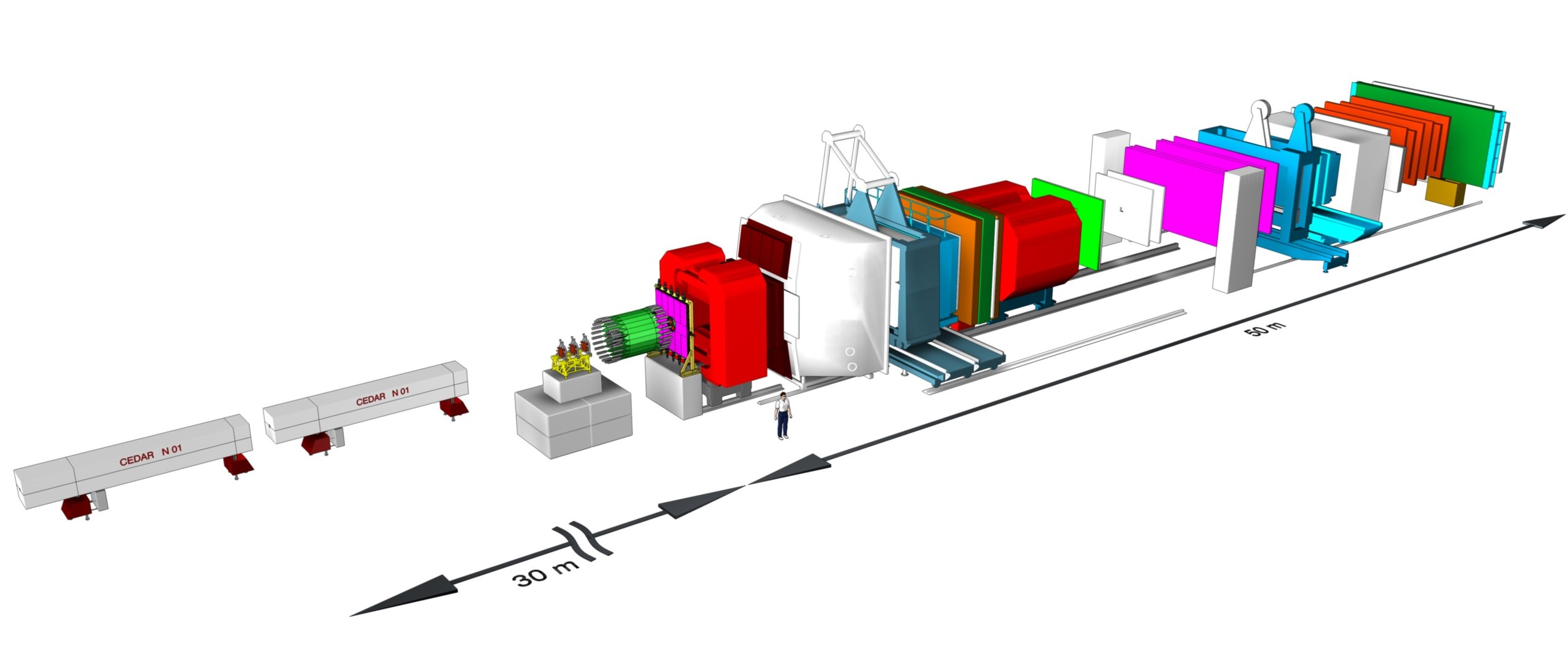}
  \caption{\em Artistic view of the COMPASS experiment}
  \label{fig:tprime2}
\end{figure}

The versatile beam line delivers secondary hadron beams with a momentum of $190\,\mathrm{GeV}/c$, a maximum intensity of $5\times10^7$ particles per second and the possibility to select both charges. The negative hadron beam consists of 96.0\% $\pi^-$, 3.5\% $K^-$ and 0.5\% $\bar p$, the while 71.5\% of the positive beam are protons, 25.5\% $\pi^+$ and 3.0\% $K^+$. Two Cherenkov Differential Counters with Achromatic Ring Focus (CEDAR) upstream of the target identify the incoming beam particles.

During a pilot run in 2004 and long subsequent data taking periods in 2008 and 2009, COMPASS has acquired large data sets of both beams impinging on liquid hydrogen (lH$_2$), nickel (Ni), tungsten (W) and lead (Pb) targets. Its large acceptance, high resolution, and high rate capability provide an excellent opportunity to study the spectrum of light mesons up to masses of $3\,\mathrm{GeV}/c^2$. The variety of production mechanisms as well as decay channels will certainly shed light into the disputed field.

\subsection{Partial-Wave Analysis}

The tool of Partial-Wave Analysis is used to disentangle the overlapping resonances and pin down the relevant quantum numbers like spin and parity. In the following, the technique will be explained on the example of a three charged pions final state.

\begin{figure}[h]
  \centering
  \includegraphics[width=.4\textwidth]{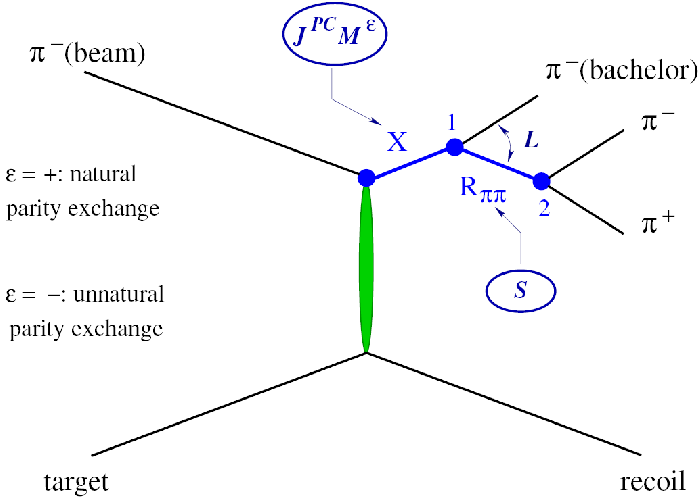}
  \caption{\em Diffractive production of a resonance $X$ ($J^{PC}M^\epsilon$) via $t$-channel Reggeon exchange and its decay into the $\pi^-\pi^+\pi^-$ final state as described in the isobar model}
  \label{fig:tprime2}
\end{figure}

The isobar model \cite{iso} is used to decompose the resonance decay into a chain of successive two-body decays as shown in Figure. $X$ with quantum numbers $I^G J^{PC}$ and spin projection $M^\epsilon$ decays into a di-pion resonance, the so called isobar, and a bachelor pion. The isobar has spin $S$ and a relative orbital momentum $L$ with respect to the pion. A partial wave is thus fully defined by $J^{PC} M^\epsilon [\textrm{isobar}] L$, where $\epsilon = \pm1$ is the reflectivity. As pomeron-exchange dominates at $190\,\mathrm{GeV}/c$, Isospin $I$ and $G$-Parity are conserved. Mostly positive reflectivity waves are used to describe the data, which corresponds to the production with natural parity exchange.

The PWA itself is performed in two steps.  First, the production amplitudes are determined in $40\,\mathrm{MeV}/c^2$ wide bins of the invariant mass $m_X$ by extended maximum likelihood fits to the acceptance corrected data. No assumption on the resonant behaviour of $X$ is made at this point, other than it's production strength stays constant within the bin width. In the example, the PWA model includes five $\pi^+\pi^-$ isobars \cite{iso2}: $(\pi\pi)_{S-wave}$, $\rho$(770), $f_0$(980), $f_2$(1270), and $\rho_3$(1690). They were described using relativistic Breit-Wigner (BW) line shape functions including Blatt-Weisskopf barrier penetration factors\cite{bw}. For the $S$-wave, we use the parametrisation from \cite{swave} with the $f_0$(980) subtracted from the elastic $\pi\pi$ amplitude and added as a separate BW resonance. The total set of partial-waves (42 in this example) also includes an isotropic background wave to take into account non-resonant production mechanisms.

Subsequently, the mass-dependence of the spin-density matrix is fitted using $\chi^2$-minimisation and a reduced set of waves. Only waves with significant intensity and phase motion in the first step are chosen. The mass dependence of the production amplitudes is parametrised by relativistic BW functions and coherent exponential background. In order to account for spin-flip and spin-non-flip amplitudes at the target vertex, the spin density matrix has a rank of two.

\section{Diffractive Dissociation}

\subsection{Observation of a $J^{PC}=1^{-+}$ exotic resonance on Pb target}

\begin{figure}[b]
  \begin{minipage}[]{.48\textwidth}
    \centering
    \includegraphics[width=.95\textwidth]{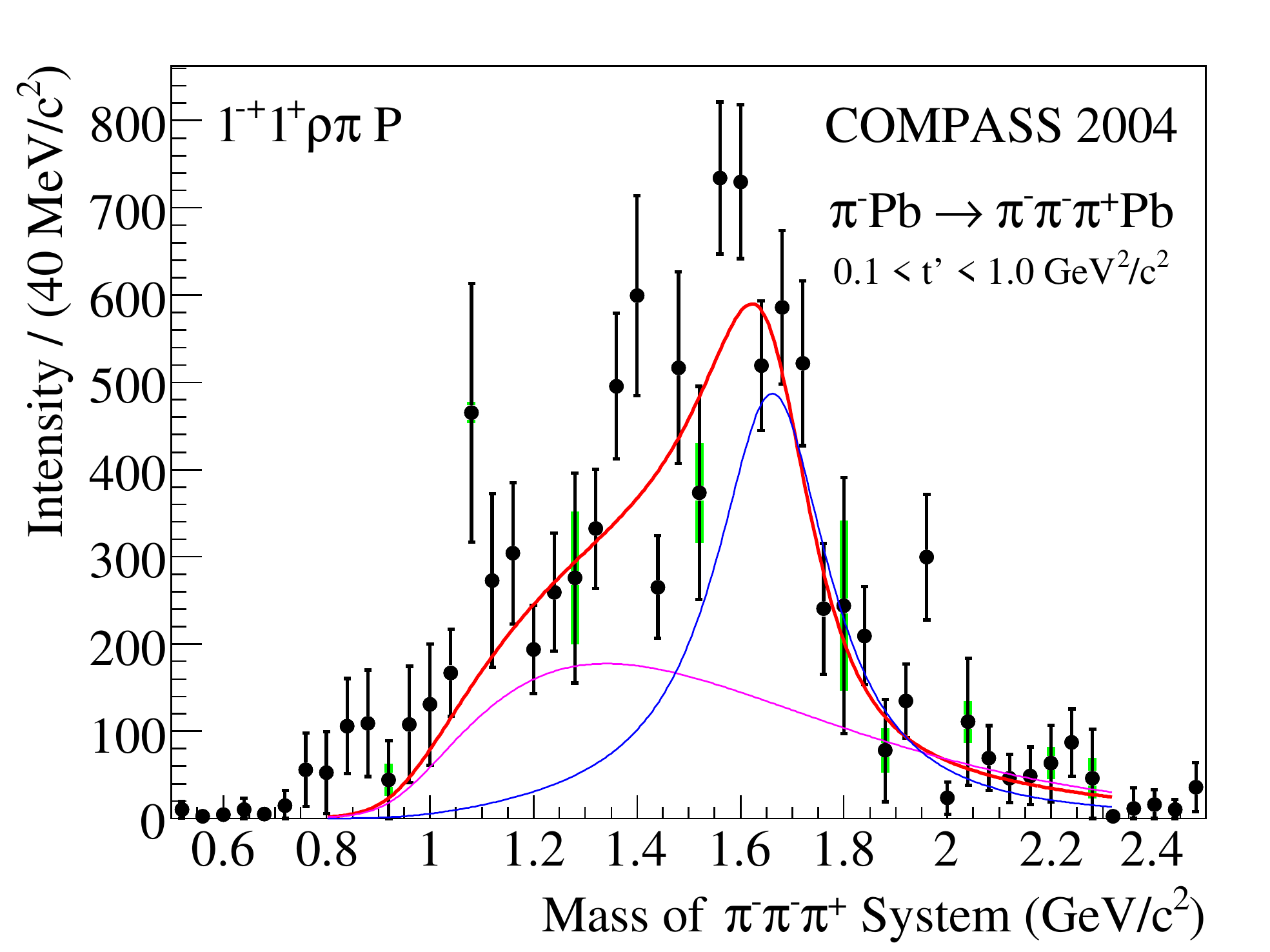}
    \caption{\em Fitted intensity of the exotic $1^{-+}1^+[\rho\pi]$ P-wave}
    \label{fig:exotic}
  \end{minipage}
  \hfill
  \begin{minipage}[]{.48\textwidth}
    \centering
    \includegraphics[width=.95\textwidth]{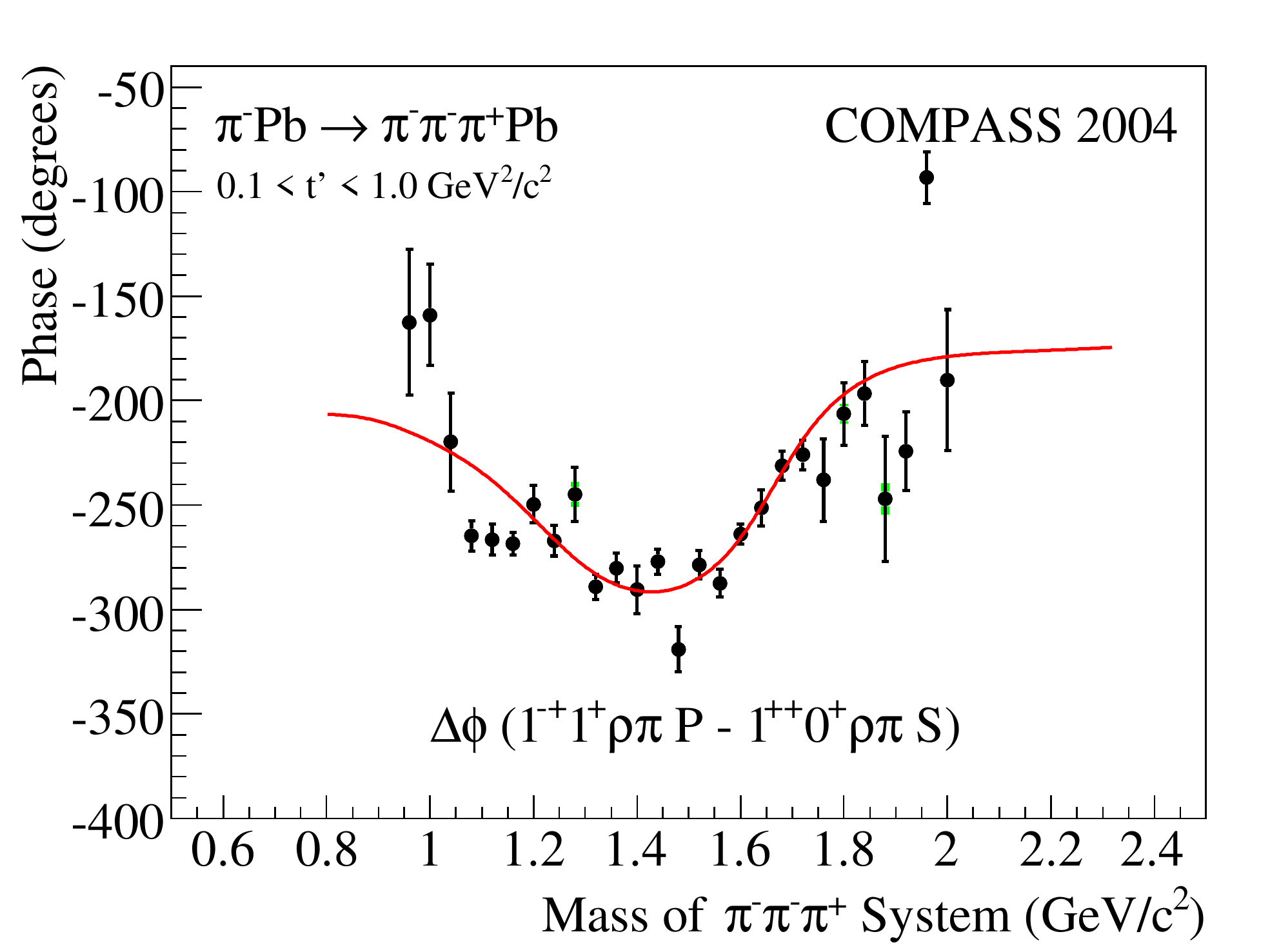}
    \caption{\em Phase-motion of the exotic $1^{-+}1^+$ wave vs. the $1^{++}0^+$ wave}
    \label{fig:phase}
  \end{minipage}
\end{figure}

Up to now, the full two-step partial-wave analysis has only been performed on the 2004 pilot run data on Pb target, which  lead to the observation of a significant spin-exotic $J^{PC} = 1^{-+}$ signal\cite{iso2}. Figure~\ref{fig:exotic} shows the intensity of this wave. The black points represent the mass-independent fit, whereas the result of the mass-dependent one is overlaid as solid line. Breit-Wigner (dashed) and background (dotted) contributions are separated. Especially the phase difference (cf.~Figure~\ref{fig:phase}) between this wave and the clean and well-established $a_2$(1260) ($1^{++}0^+\rho\pi$ S-wave) clearly indicates the resonant nature. Both the mass of $1660 \pm 10^{+0}_{-64}\,\mathrm{MeV}/c^2$ and the width of $269 \pm 21^{+42}_{-64}\,\mathrm{MeV}/c^2$ are consistent with the controversial $\pi_1$(1600).

\subsection{3$\pi$ final states on lH$_2$ target}

\begin{figure}[]
  \subfigure[]{\includegraphics[width=.48\textwidth, height=.19\textheight]{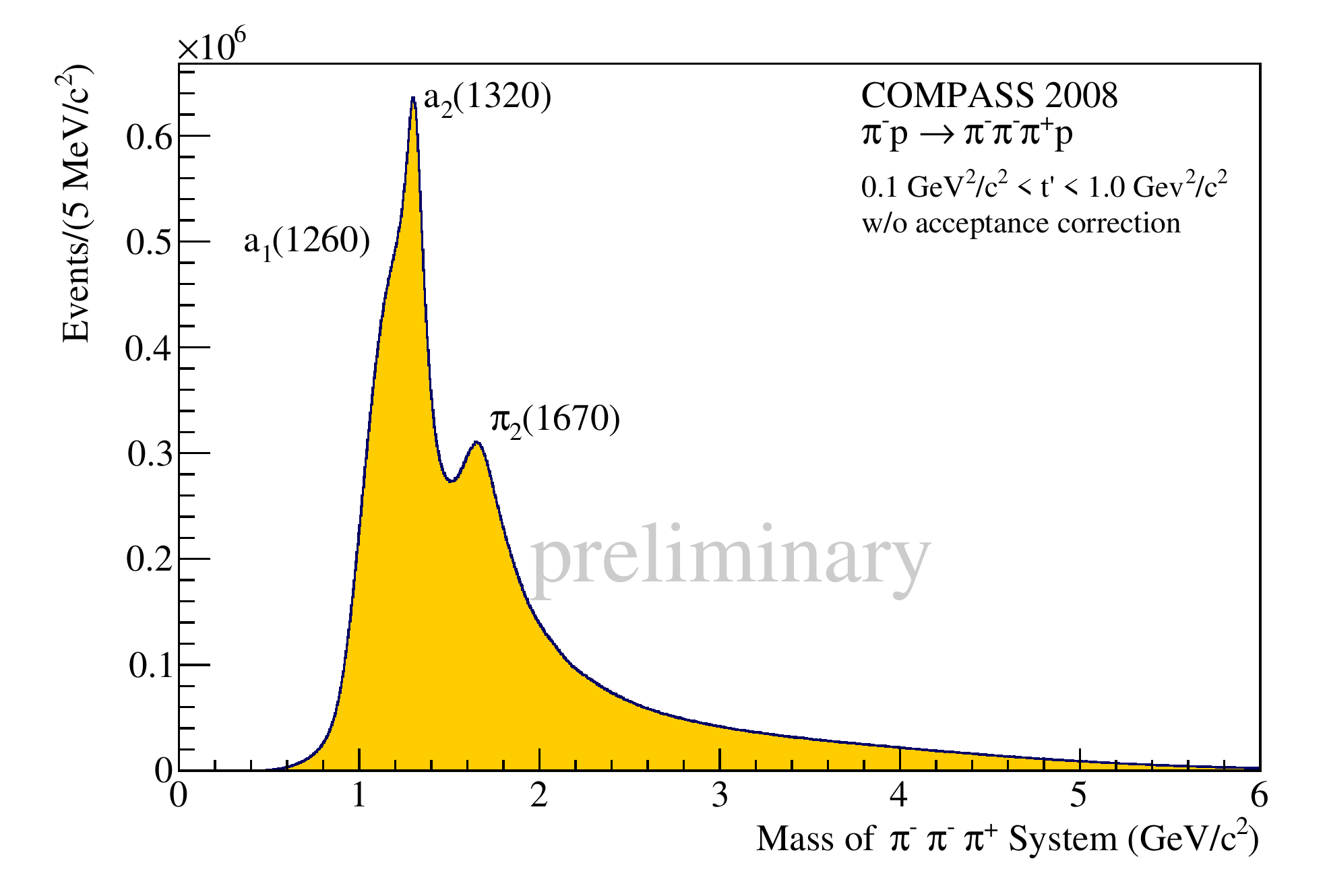}}
  \subfigure[]{\includegraphics[angle=90, width=.48\textwidth, height=.19\textheight]{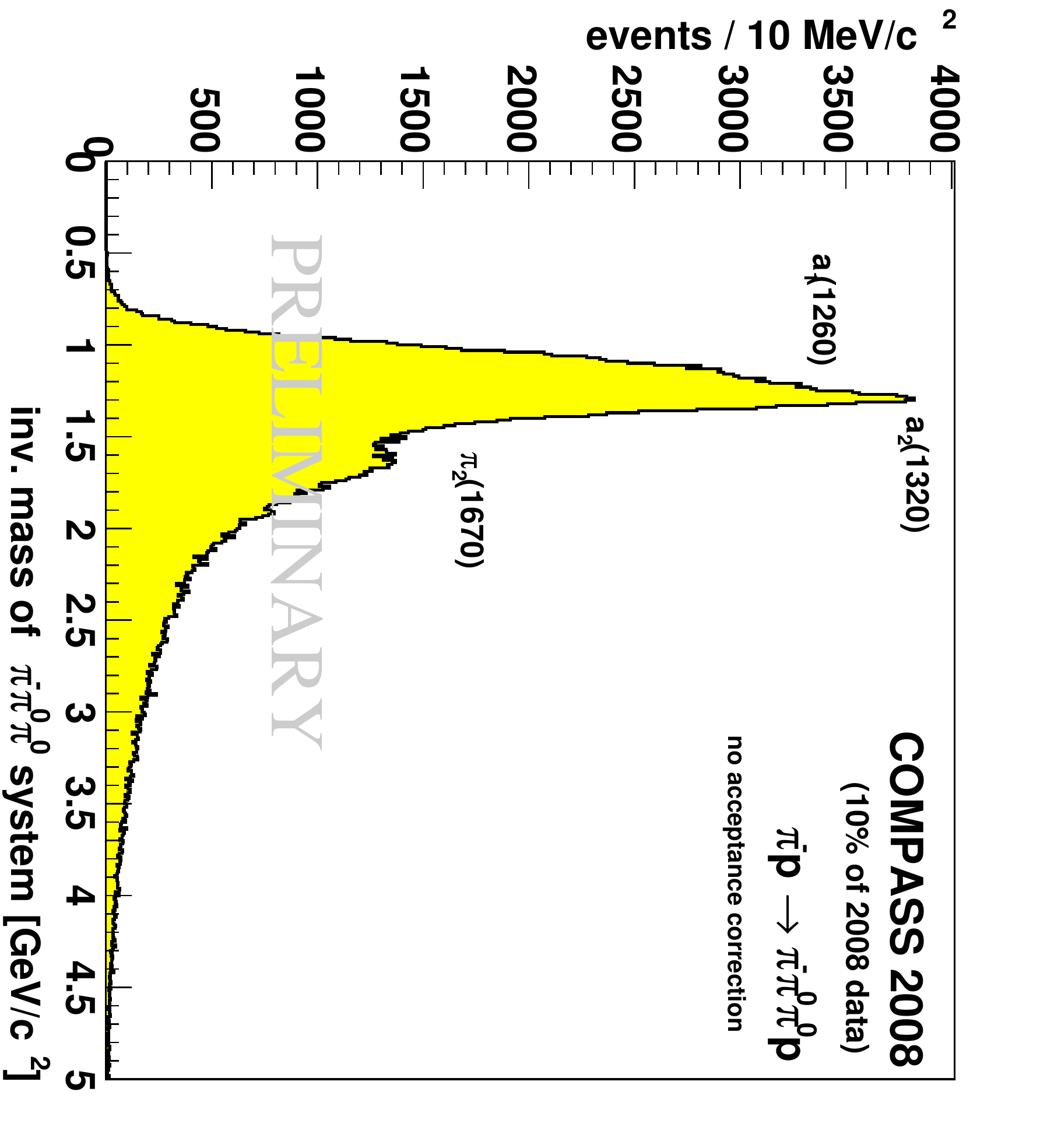}}
  \subfigure[]{\hspace{1cm}\includegraphics[width=.35\textwidth]{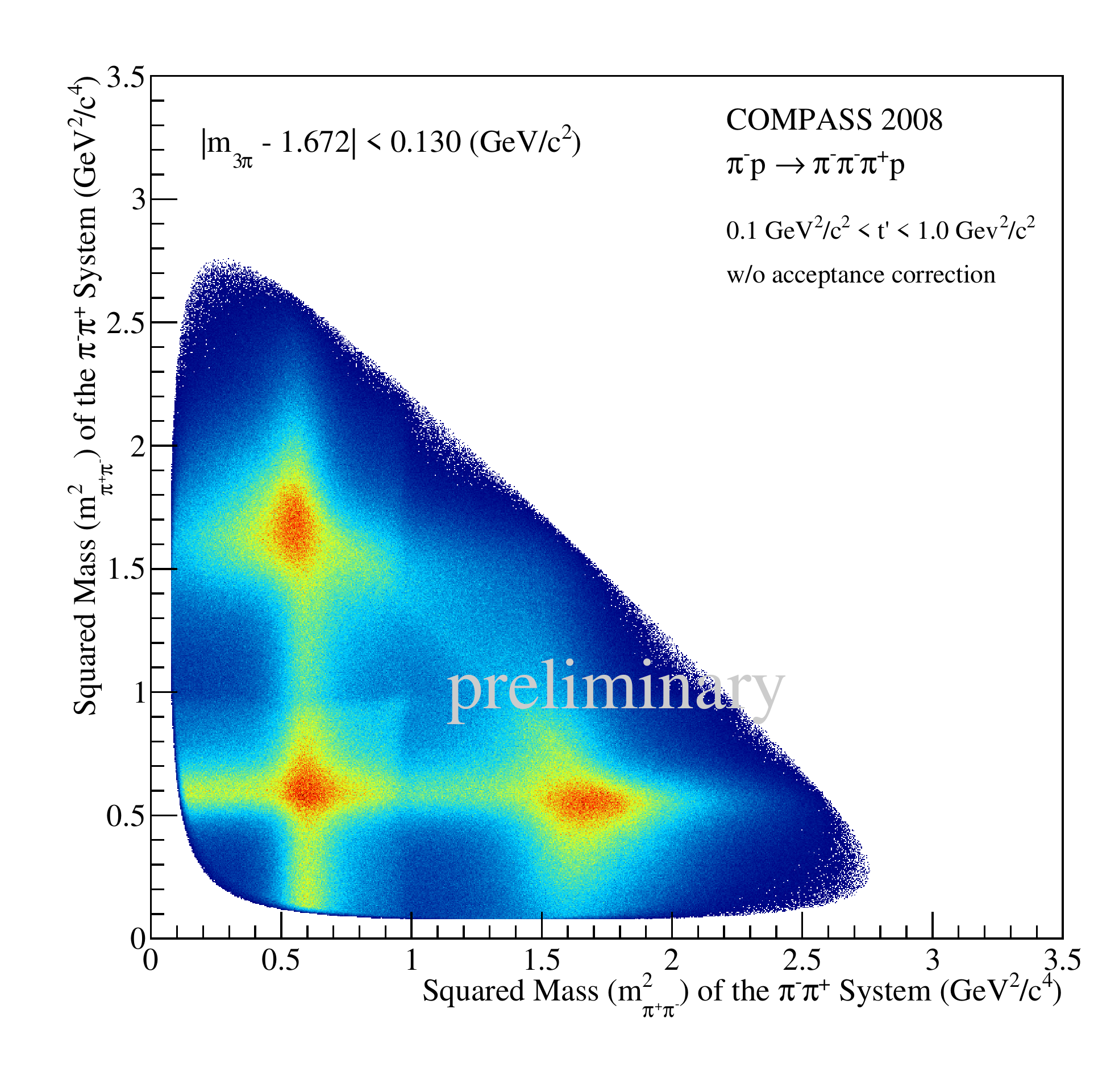}\hspace{1cm}}
  \subfigure[]{\includegraphics[width=.46\textwidth, height=.18\textheight]{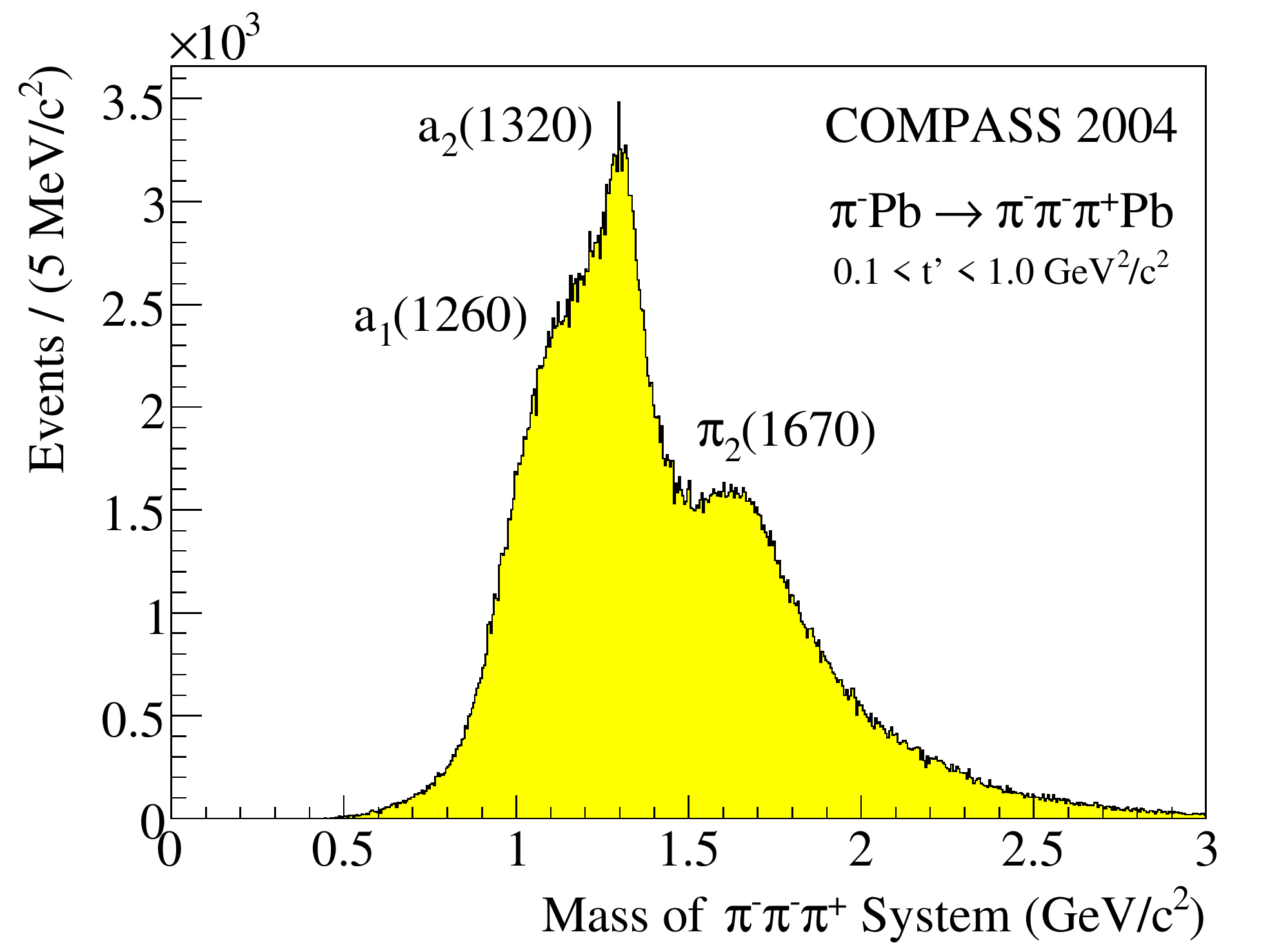}}
  \caption{{\em Invariant mass distribution for} (a) {\em $\pi^-\pi^+\pi^-$} (b) {\em $\pi^-\pi^0\pi^0$, both for the 2008 data sample on lH$_2$ target} (c) Dalitz distribution of the $\pi_2$(1670) ($\pm 0.5\Gamma$) (d) {\em $\pi^-\pi^+\pi^-$ of the 2004 data sample on Pb for comparison}}
  \label{fig:invmass}
\end{figure}

With the improved COMPASS spectrometer in 2008, the $\rho\pi$ decay channel of the $\pi_1$(1600) has been studied in two modes of the $3\pi$ final state, $\pi^-\pi^+\pi^-$ (charged) and $\pi^-\pi^0\pi^0$ (neutral), both of them with unprecedented statistics (cf.~Figure~\ref{fig:invmass}). The Dalitz plot for the $\pi_2$(1670) resonance in Figure~\ref{fig:invmass}(c) illustrates as an example the quantity and the resolution of the data.

\begin{figure}[h]
  \begin{minipage}[]{.48\textwidth}
    \centering
    \includegraphics[width=.95\textwidth]{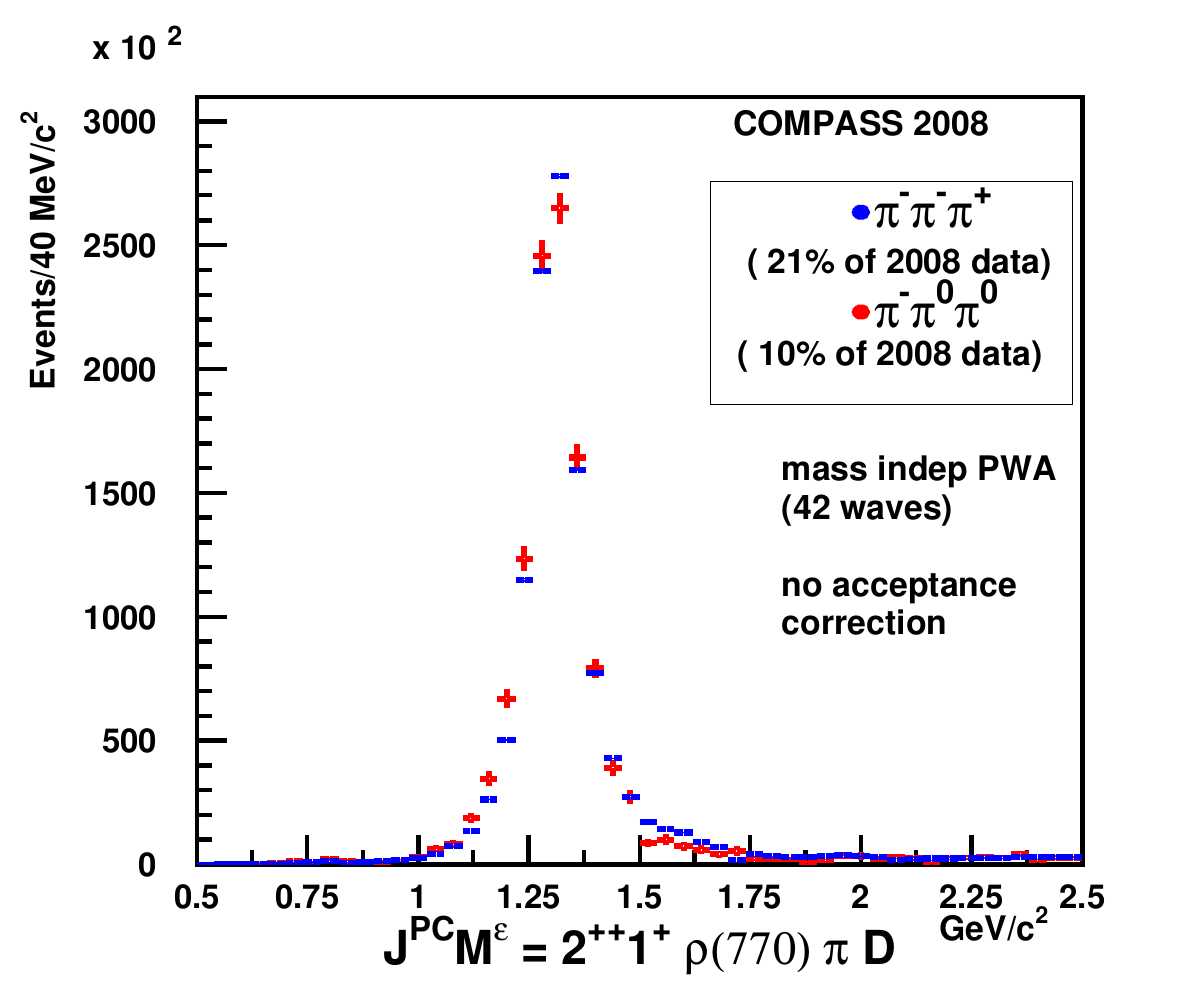}
    \includegraphics[width=.95\textwidth]{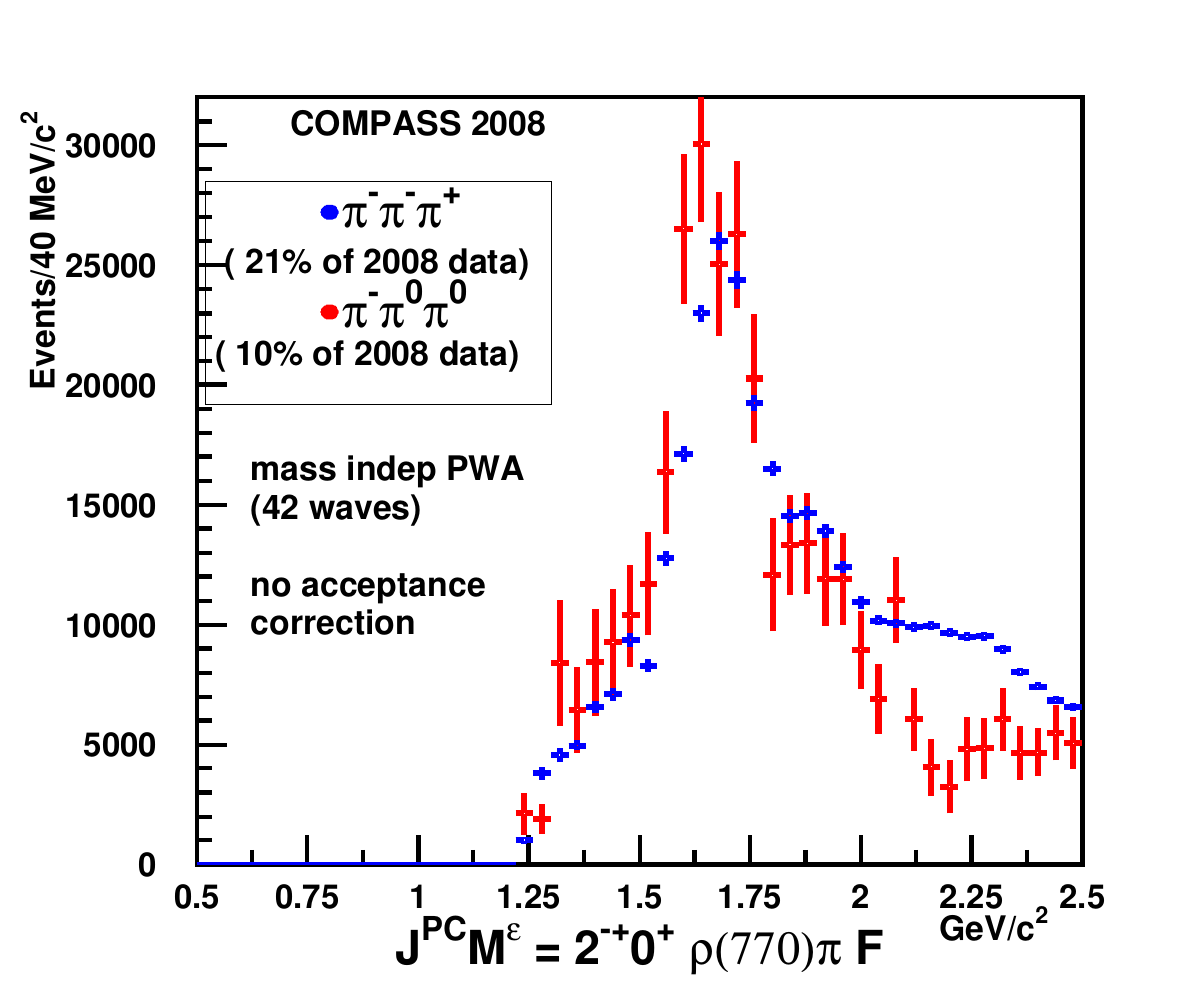}
  \end{minipage}
  \begin{minipage}[]{.48\textwidth}
    \centering
    \includegraphics[width=.95\textwidth]{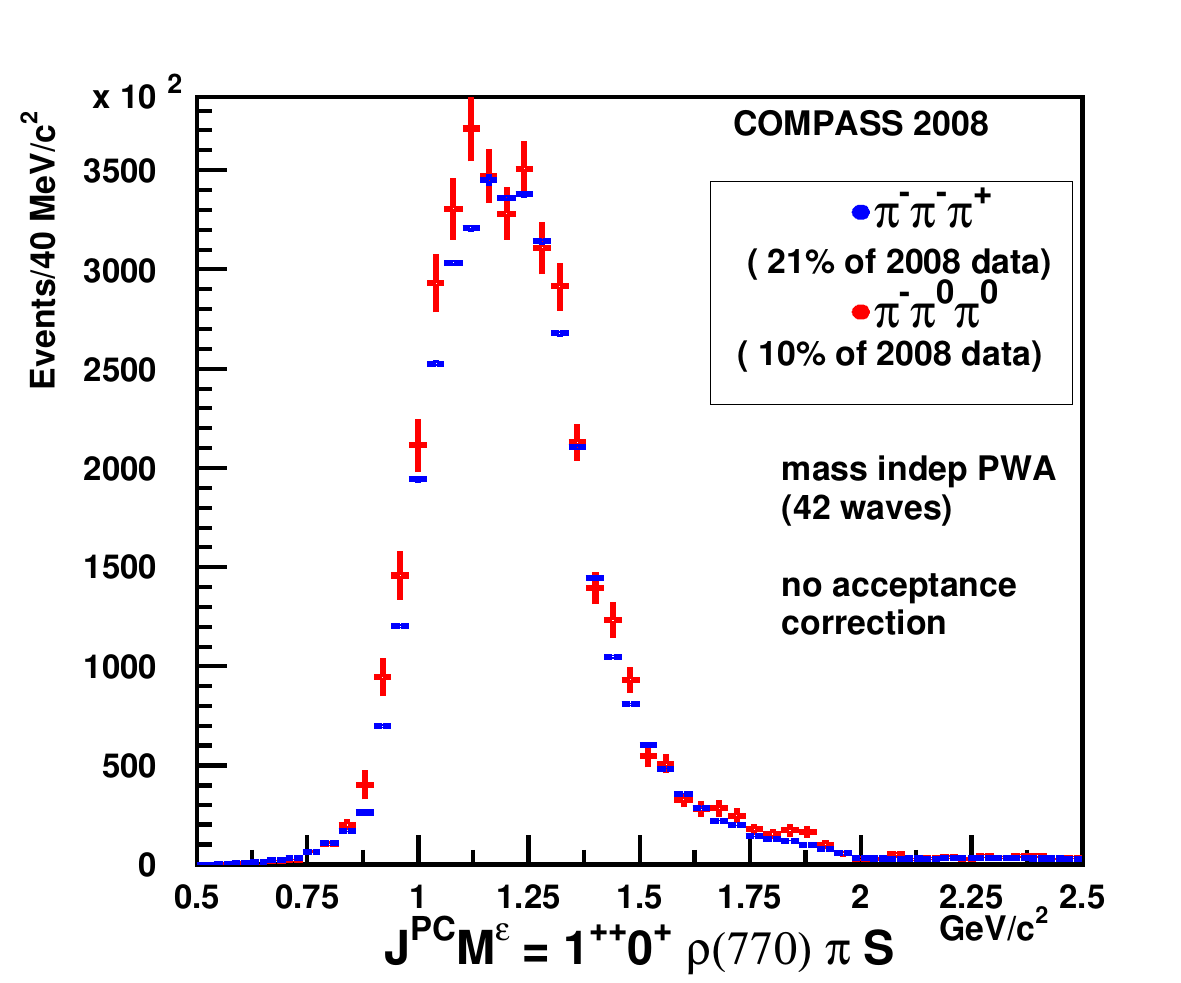}
    \includegraphics[width=.95\textwidth]{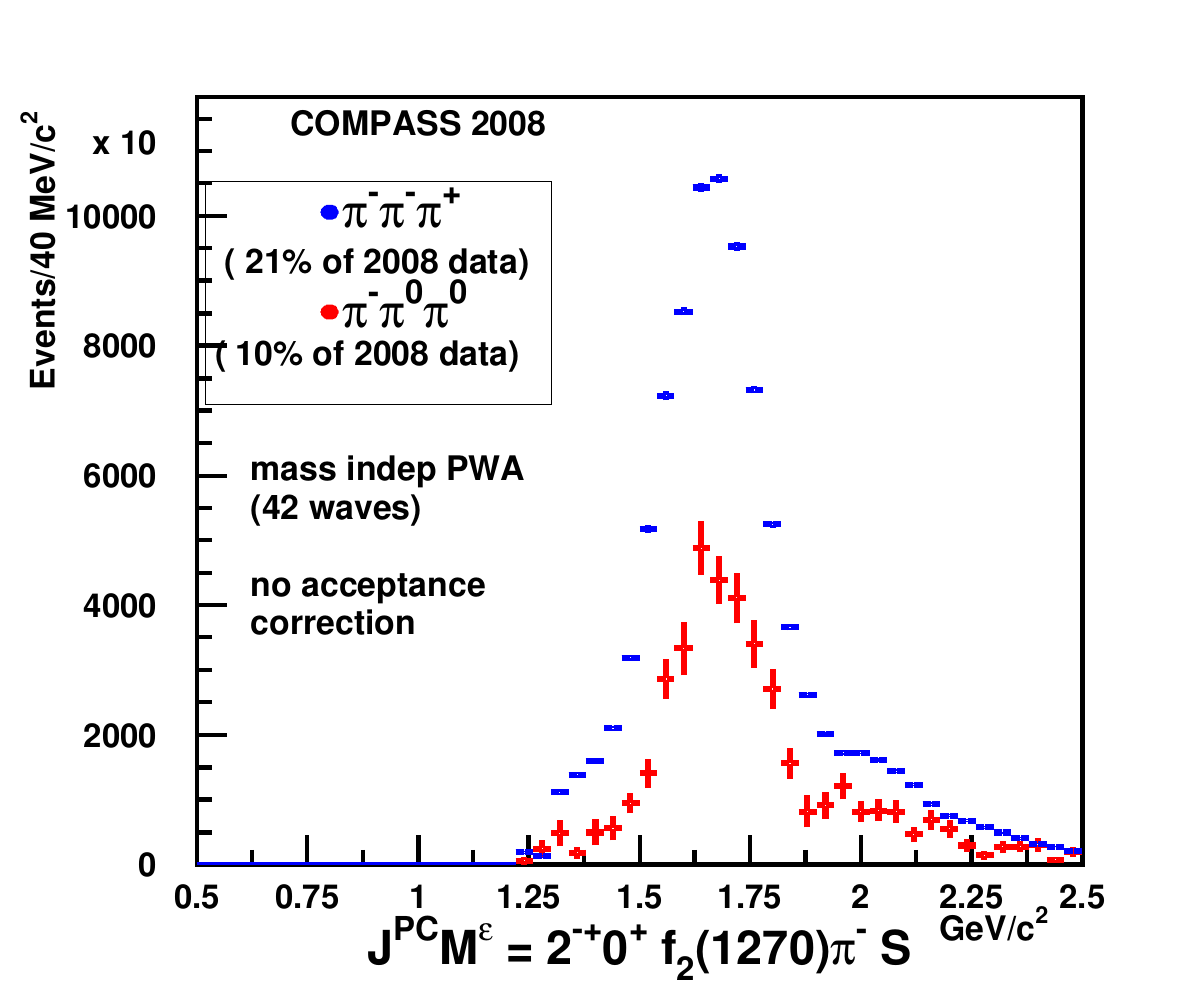}
  \end{minipage}
  \caption{\em Comparison of PWA intensities of main waves for neutral vs. charged mode. (Top/Left) Intensities of the $a_2$ ($2^{++}1^{+}\rho\pi$ D-wave) used for normalisation. (Top/Right) $a_1$ ($1^{++}0^+\rho\pi$ D-wave) (Bottom Left) $\pi_2$ ($2^{-+}0^+f_2\textrm{(1270)}$ S-wave) (Bottom Right) $\pi_2$ ($2^{-+}0^+\rho\pi$ F-wave)}
  \label{fig:isospin}
\end{figure}

A mass-independent partial-wave analysis has been performed for both modes, applying the same model as for the 2004 result. The data set has not yet been corrected for acceptance, though experience from the pilot run data does not show a large dependence on most kinematic variables. The wave intensities shown in Figure~\ref{fig:isospin} are normalised to the clean $a_2$(1320) ($2^{++}1^+$ D-wave) signal to compensate for the different reconstruction efficiencies and the not-yet-corrected acceptances. Similar intensities for the $\rho\pi$ decay can be found, whereas a suppression factor of approximately two is observed for the waves decaying into $f_2\pi$. This result can be derived by a simple calculation using the isospin Clebsch-Gordan coefficients, and provides an important confirmation for all analyses availing of the greatly improved electromagnetic calorimetry. In particular, further ongoing analyses to search for exotic mesons cover prominent $\eta\pi$, $\eta'\pi$, $f_1\pi$ and $b_1\pi$ final states, where COMPASS has recorded a significantly larger data sample than previous experiments.

\clearpage

\subsection{Nuclear Effect in Meson Production}

\begin{figure}[h]
  \begin{minipage}[]{.48\textwidth}
    \centering
    \includegraphics[width=.95\textwidth]{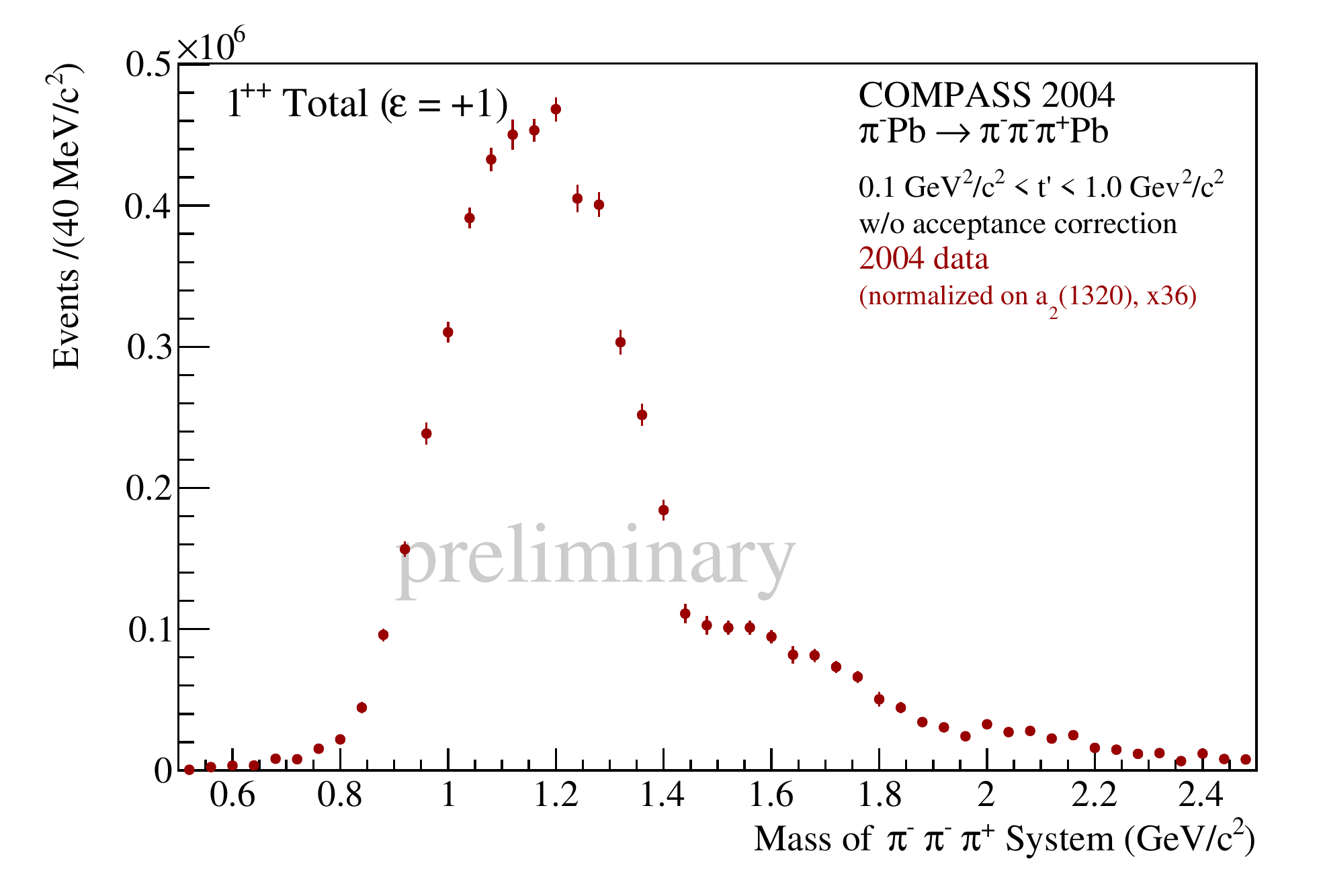}
    \includegraphics[width=.95\textwidth]{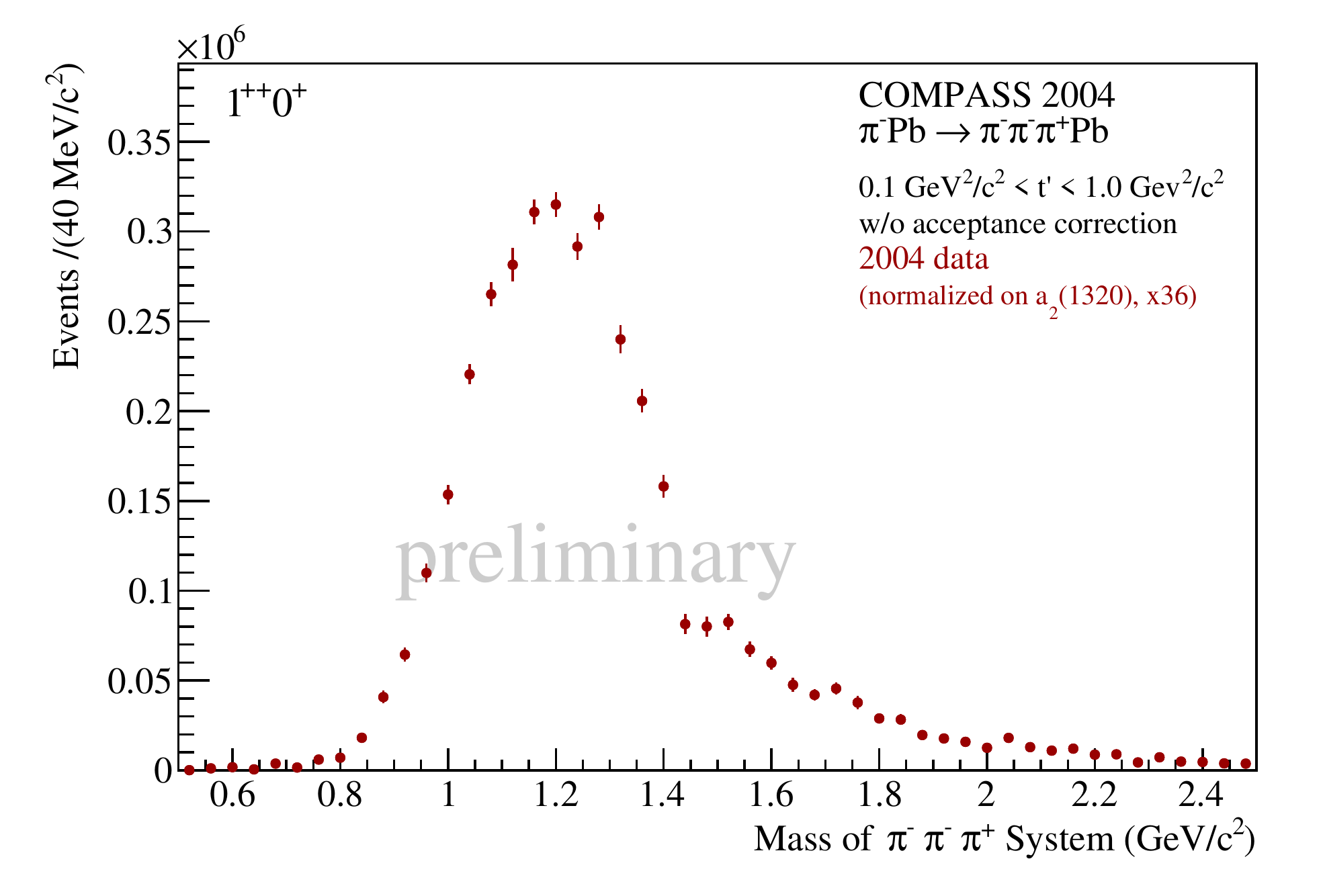}
    \includegraphics[width=.95\textwidth]{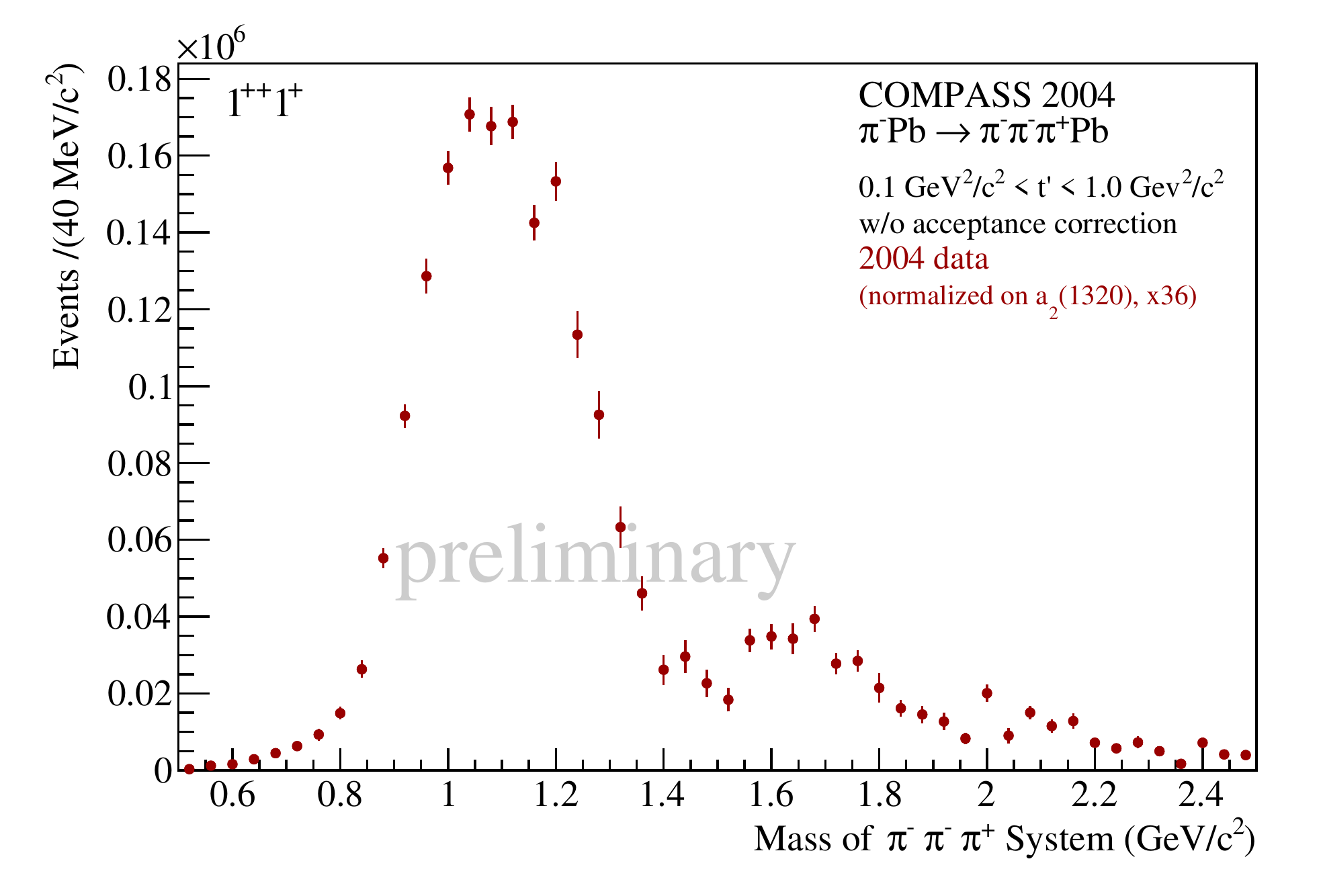}
  \end{minipage}
  \begin{minipage}[]{.48\textwidth}
    \centering
    \includegraphics[width=.95\textwidth]{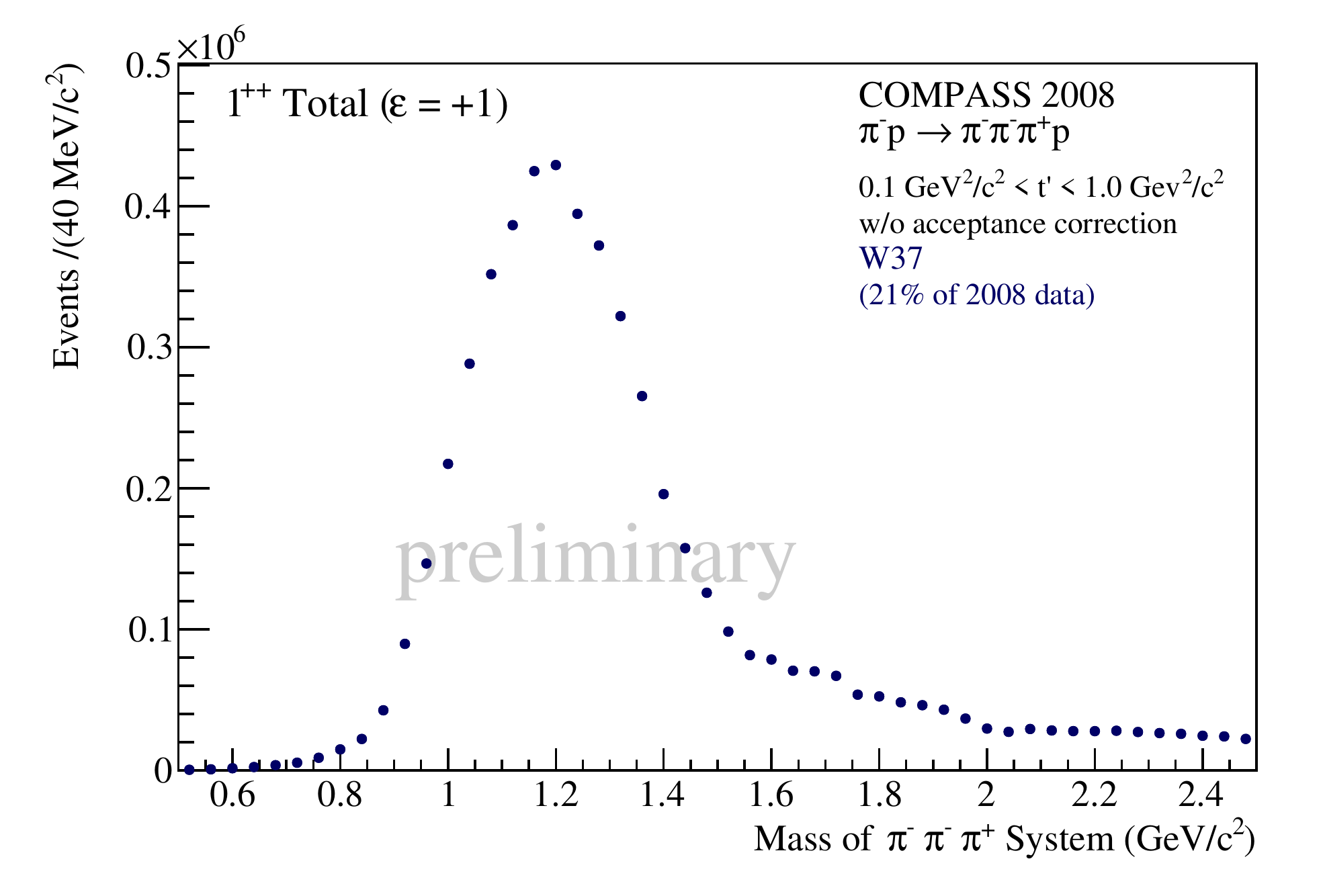}
    \includegraphics[width=.95\textwidth]{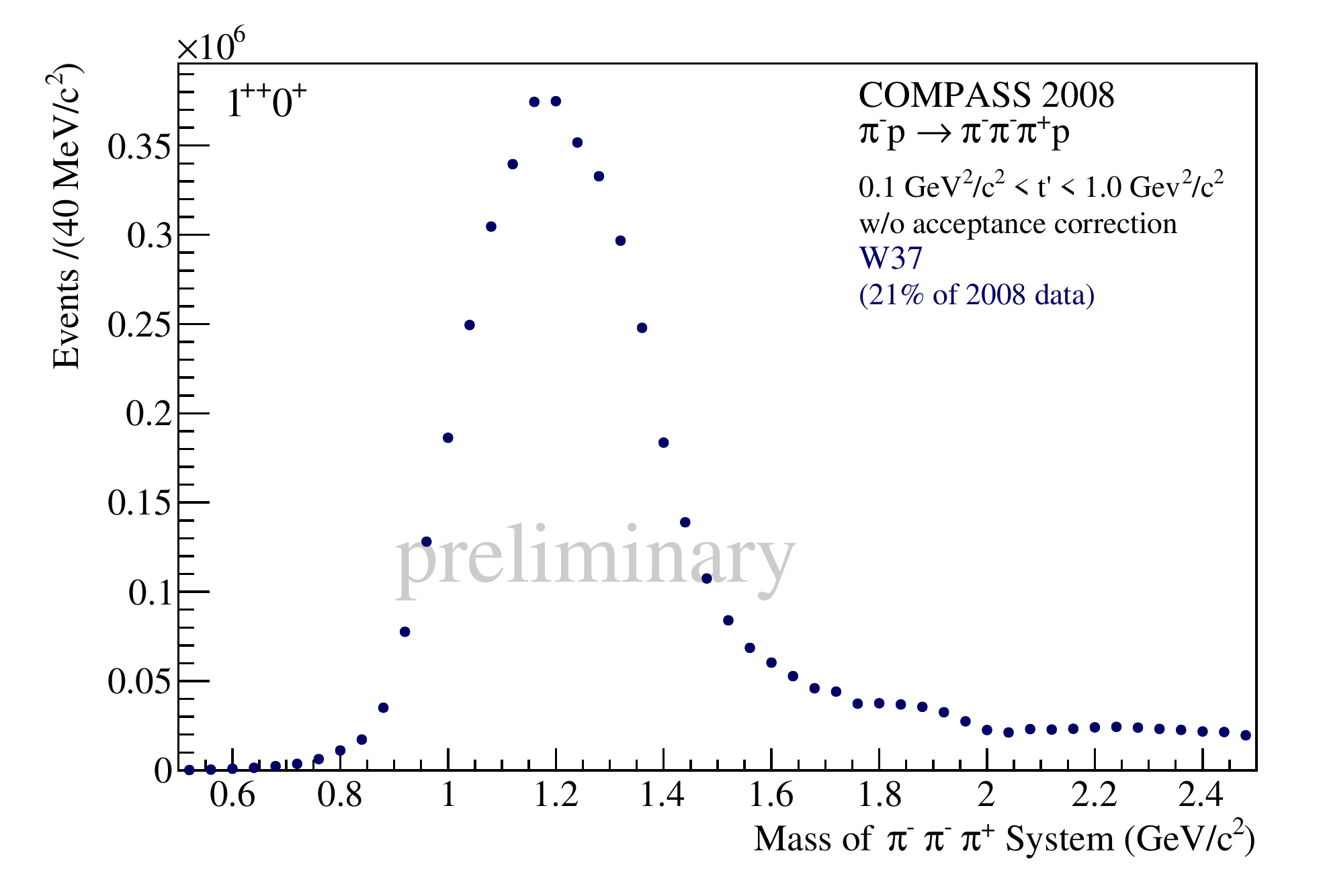}
    \includegraphics[width=.95\textwidth]{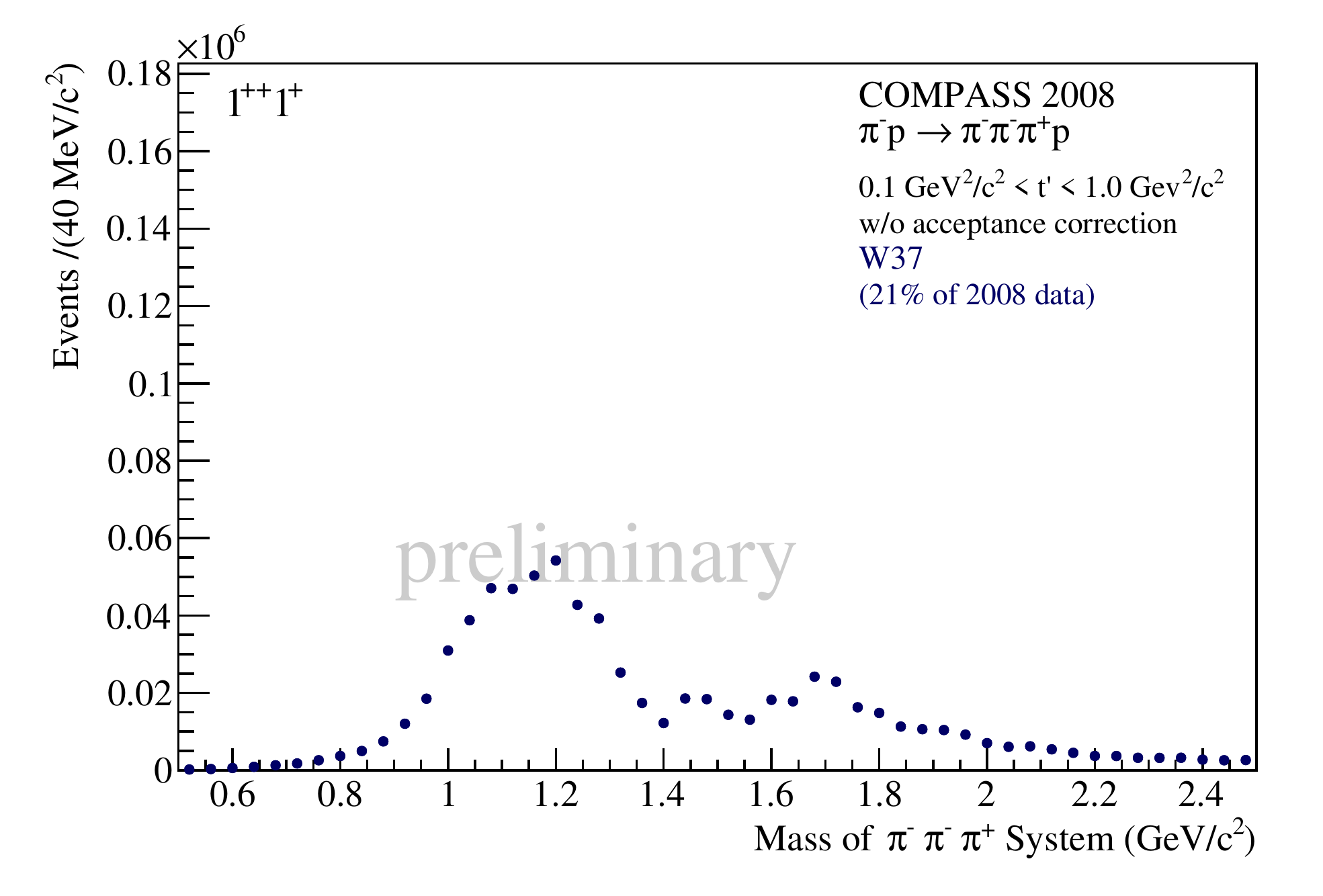}
  \end{minipage}
  \caption{\em Normalised intensity sums of the $J^{PC} = 1^{++}$ partial-waves for spin projection quantum numbers $M = 0$ (middle) and $M = 1$ (bottom). The sum over all $M$ is shown in the top row. The left column shows 2004 data on Pb target, the right column 2008 data on H$_2$ target. The wave intensities are dominated by a broad structure around $1.2\,\mathrm{GeV}/c^2$ which is the $a_1$(1260).}
  \label{fig:nuceff}
\end{figure}

A similar comparison of the data recorded with lH$_2$ and the one on lead target shows a surprisingly large dependence of the target material, given the fact that the squared four-momentum transfer distribution ($0.1 < t' < 1.0 (\mathrm{GeV}/c^2)^2$) clearly suggests nucleons as interaction partners. Using the same normalisation to $a_2$(1320) as in the previous section, the lH$_2$ data exhibit a strong suppression of $M = 1$ waves whereas the corresponding $M = 0$ waves are enhanced such that the intensity sum over $M$ remains approximately the same for both target materials. Figure~\ref{fig:nuceff} shows this effect for the $a_1$(1260) peak in the $J^{PC} = 1^{++}$ waves. Another data set recorded with a Ni target suggests a dependence on the atomic number $A$, but further investigation is needed.

\newpage

\subsection{Interference with Photo-Production}

\begin{figure}[h]
  \begin{minipage}[]{.48\textwidth}
    \centering
    \includegraphics[width=.95\textwidth]{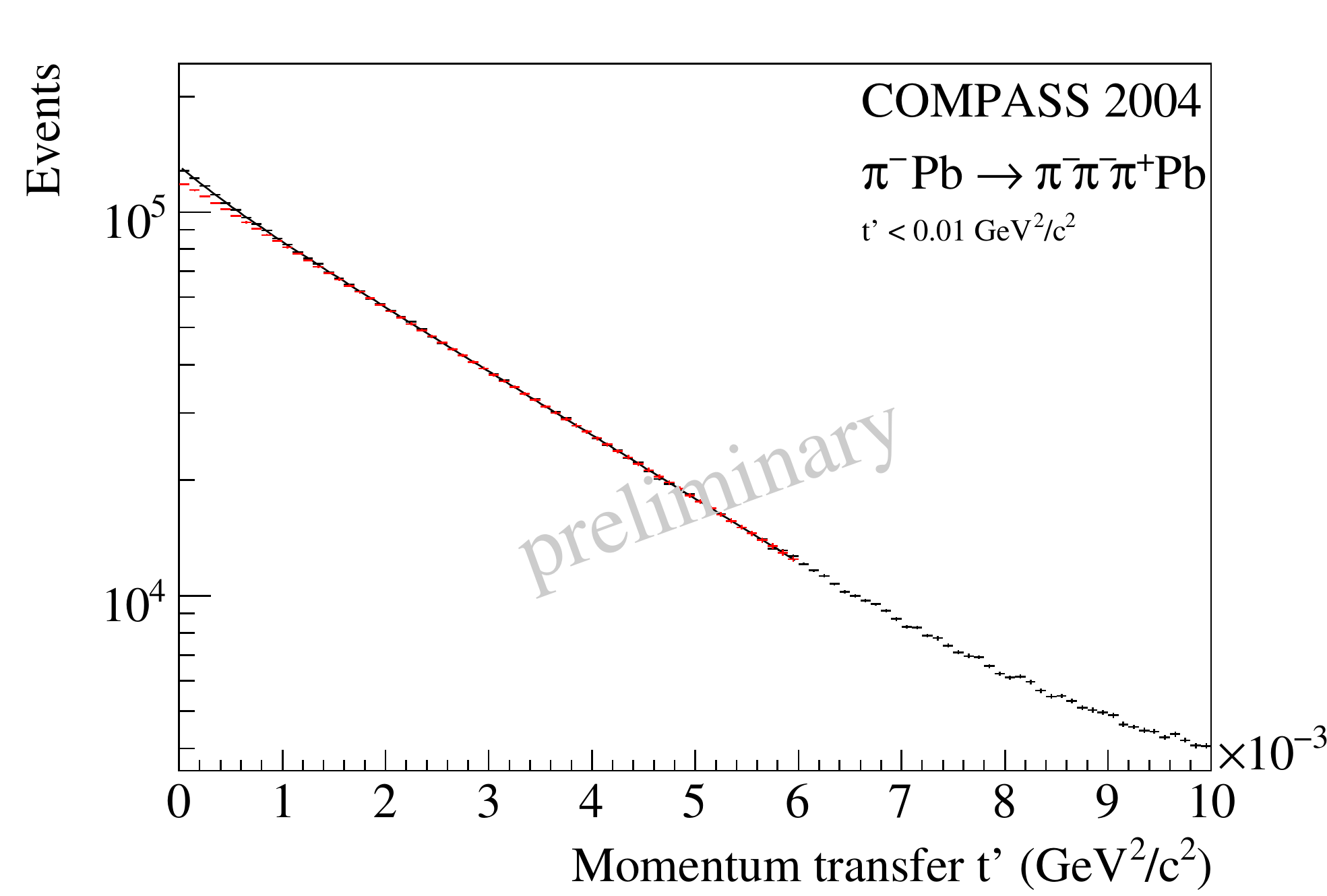}
    \includegraphics[width=.95\textwidth]{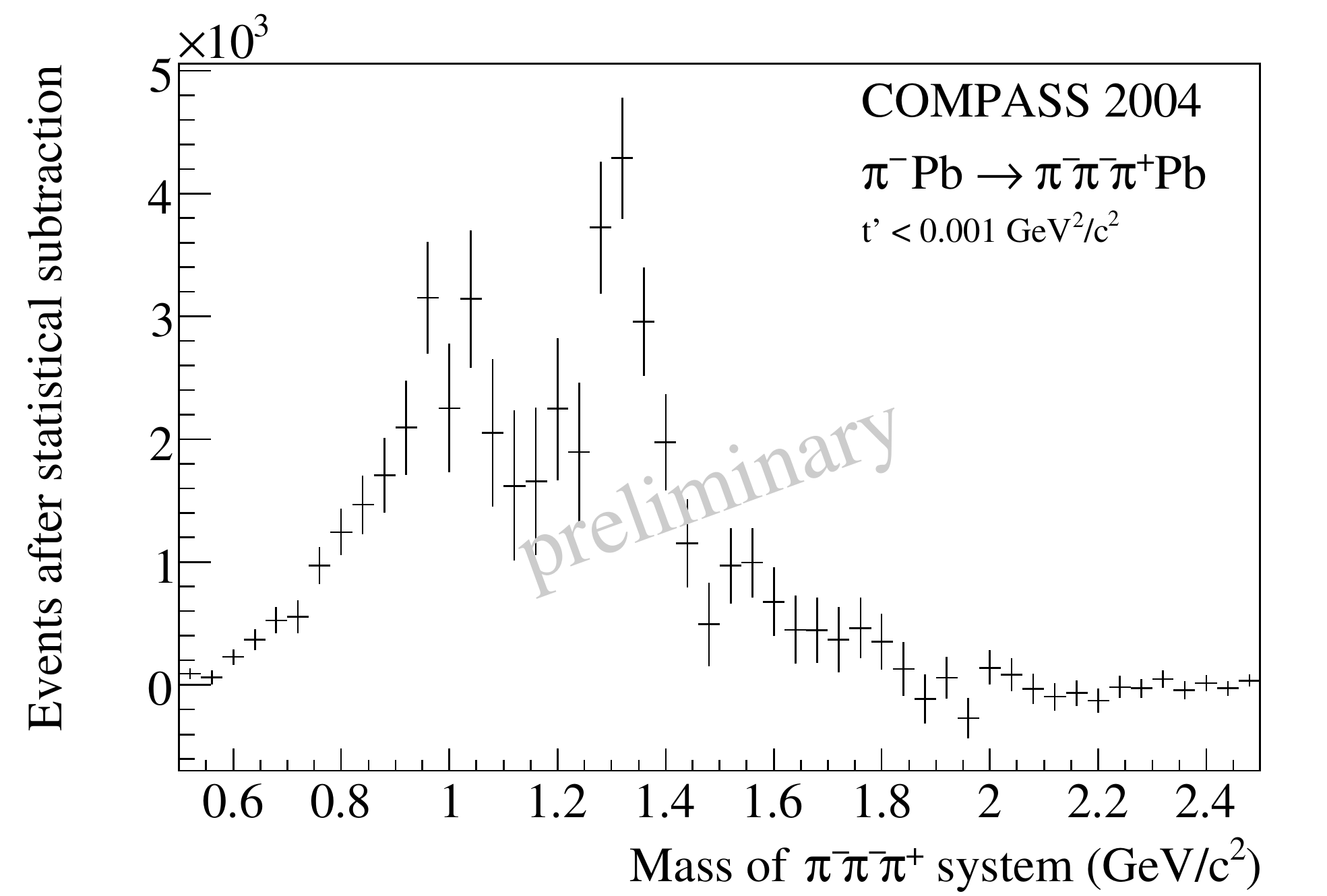}
  \end{minipage}
  \begin{minipage}[]{.48\textwidth}
    \centering
    \includegraphics[width=.95\textwidth]{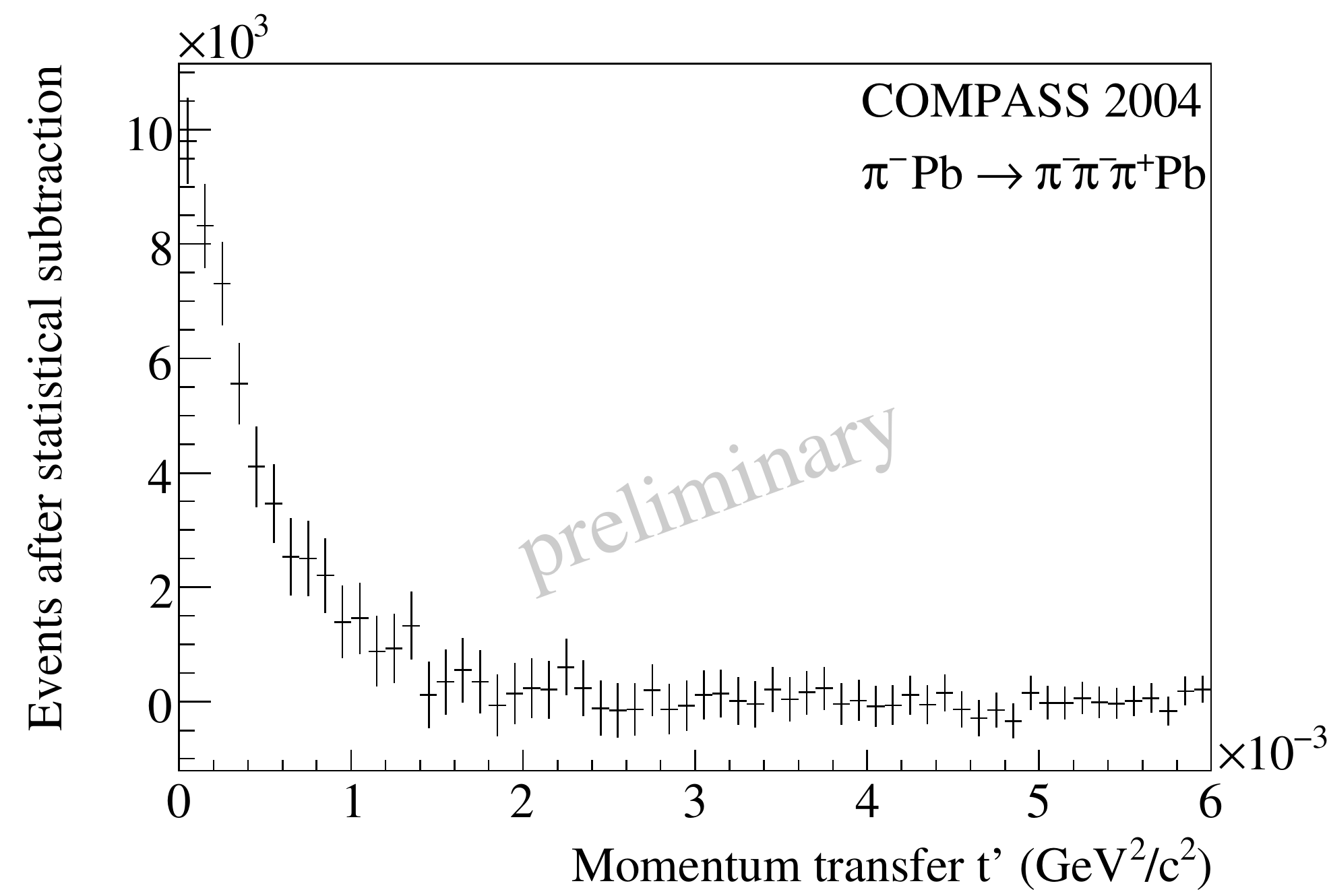}
    \includegraphics[width=.95\textwidth]{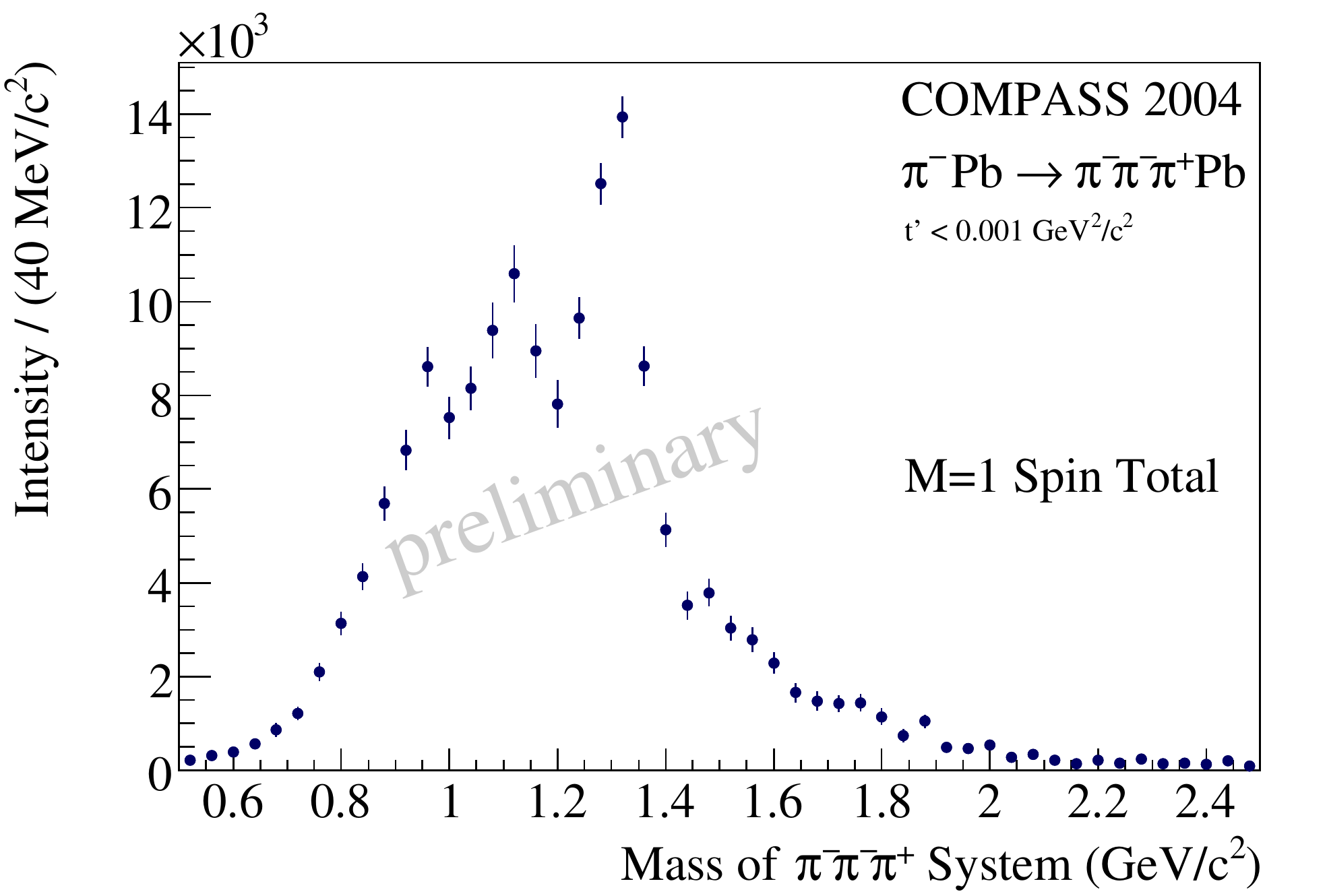}
  \end{minipage}
  \caption{\em (Top Left) Additional contribution to $t'$ spectrum: photo-production (Top Right) $t'$ spectrum after subtraction of diffractive contribution (Bottom Left) Invariant mass spectrum obtained by statistical subtraction (Bottom Right) Sum of $M = 1$ partial-waves where diffractive contribution is negligible}
  \label{fig:photo}
\end{figure}

Photon-induced reactions are studied to determine radiative widths of resonances. The presence of photo-produced events is visible as an additional contribution to the $t'$ spectrum at $t' < 0.001 (\mathrm{GeV}/c^2)^2$, where a steeper slope is observed. By fitting two exponential curves to the spectrum, one can extract the contribution of the Primakoff effect to the data at $t' \approx 0$.  Doing this statistical subtraction in bins of the $3\pi$ invariant mass results in a spectrum with a clear signature of a photo-produced $a_2$(1320) meson (cf.~Figure~\ref{fig:photo}). Equivalent mass spectra have been obtained by the Selex Collaboration\cite{selex}.

The momentum-transfer region $t'<10^{-3}\,(\,GeV/c^2)^2$ was also fitted using the PWA model. The photo-production is dominant for waves with spin projection $M=1$, because the diffractive production strength is proportional to $(t')^M\,\exp{(-bt')}$. In the bottom row of Figure~\ref{fig:photo}, the similarity of the sum of all partial waves with $M=1$ and the mass spectrum obtained by statistical subtraction in striking. The transition region between the two production mechanisms of the $a_2$(1320) meson has been subject to further studies\cite{jan}, testing inter alia recent Chiral Perturbation Theory calculations close to threshold\cite{chipt}.

\newpage

\section{Strangeness}

The possibility for particle identification of the beam as well as the final state particles provides COMPASS also with the tools to study strange meson production and diffraction, which are interesting for glueball search at central rapidities as well as diffractive production of hybrids and other not confirmed resonances.

\begin{figure}[h]
  \begin{minipage}[]{.48\textwidth}
    \centering
    \includegraphics[width=.95\textwidth]{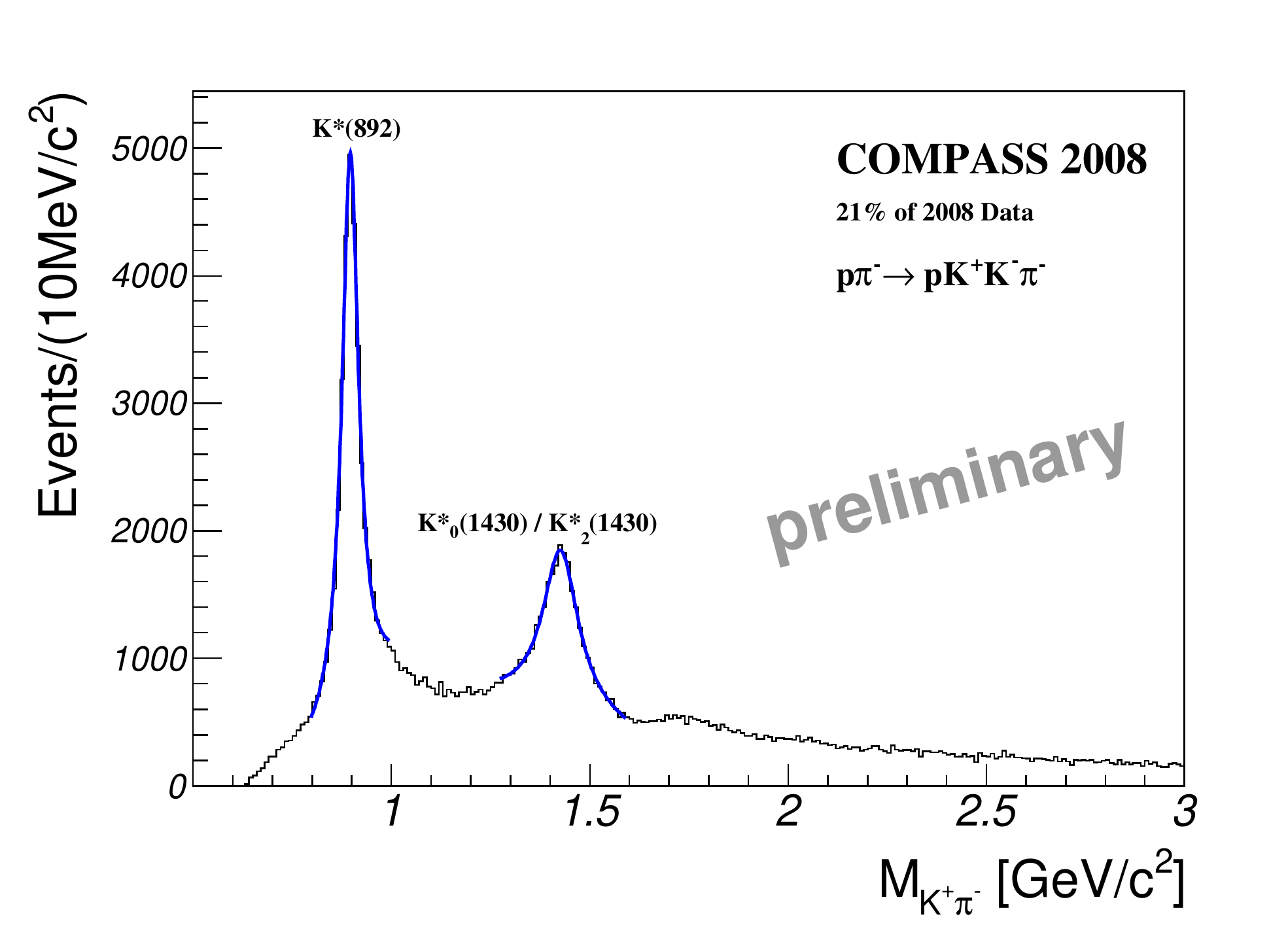}
    \includegraphics[width=.95\textwidth]{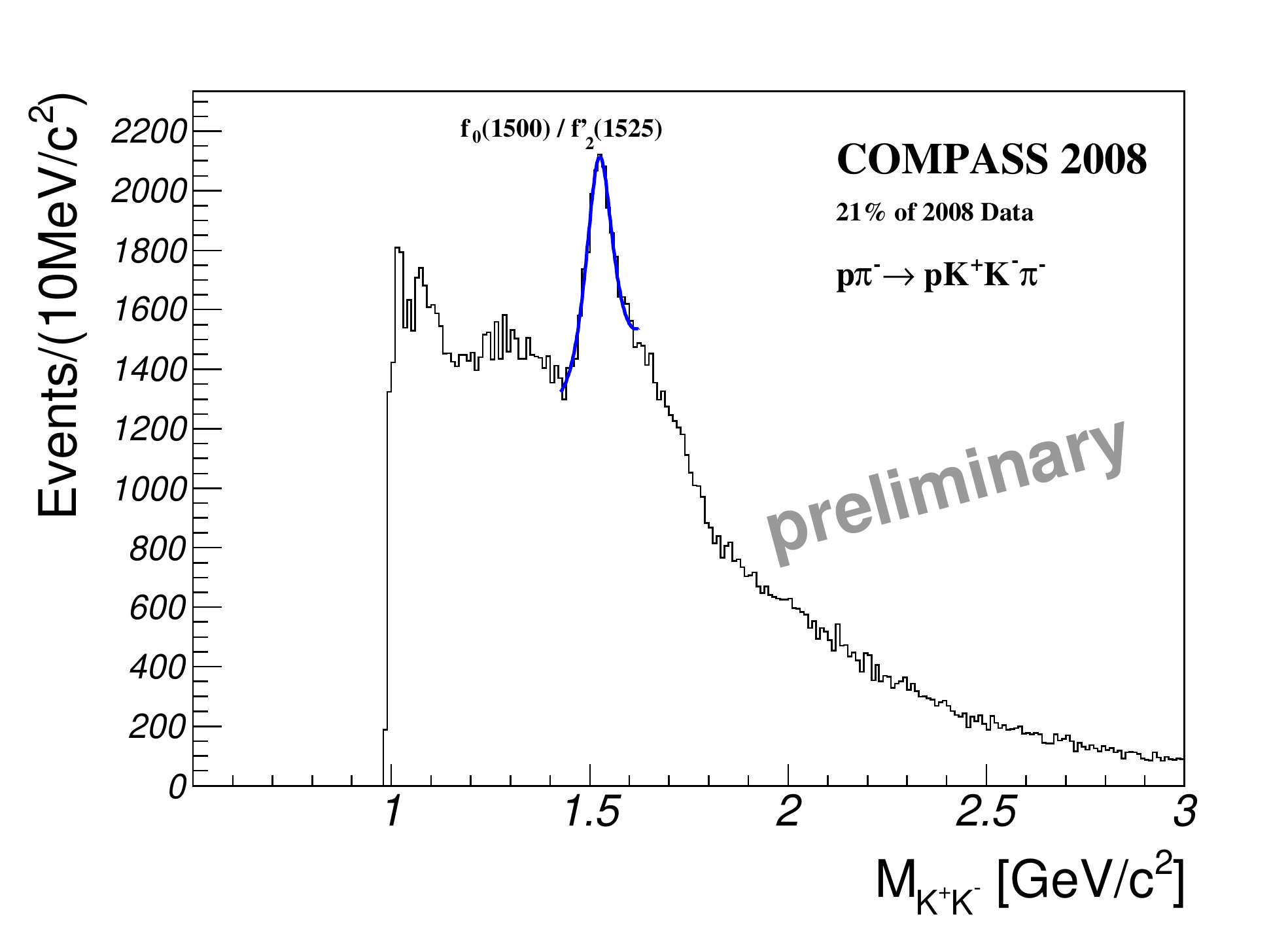}
    \caption{\em Invariant mass distribution of (Top) $\pi^-K^+$ and (Bottom) $K^+K^-$ sub-systems}
    \label{fig:Kcharged}
  \end{minipage}
  \hfill
  \begin{minipage}[]{.48\textwidth}
    \centering
    \includegraphics[width=\textwidth]{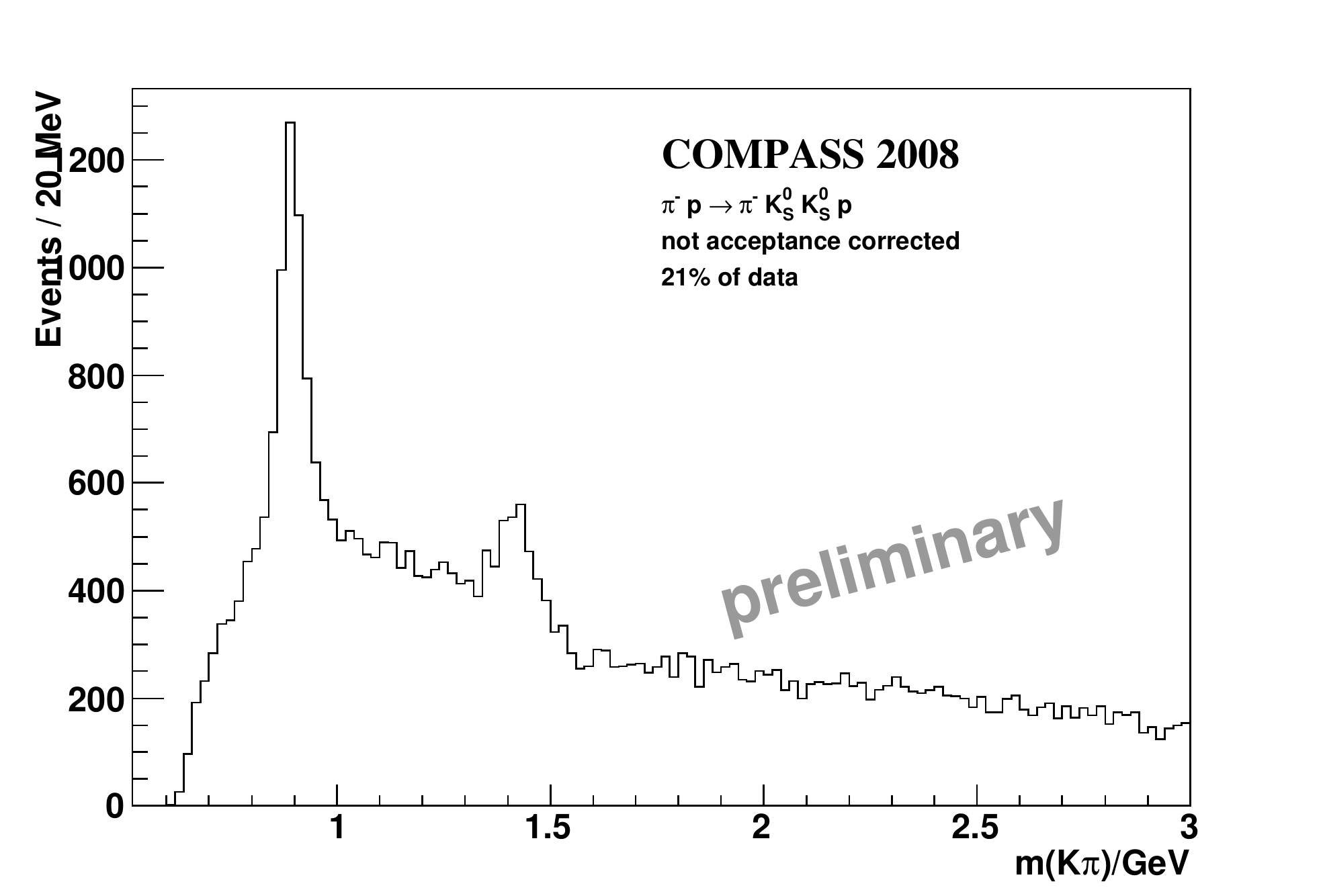}
    \includegraphics[width=\textwidth]{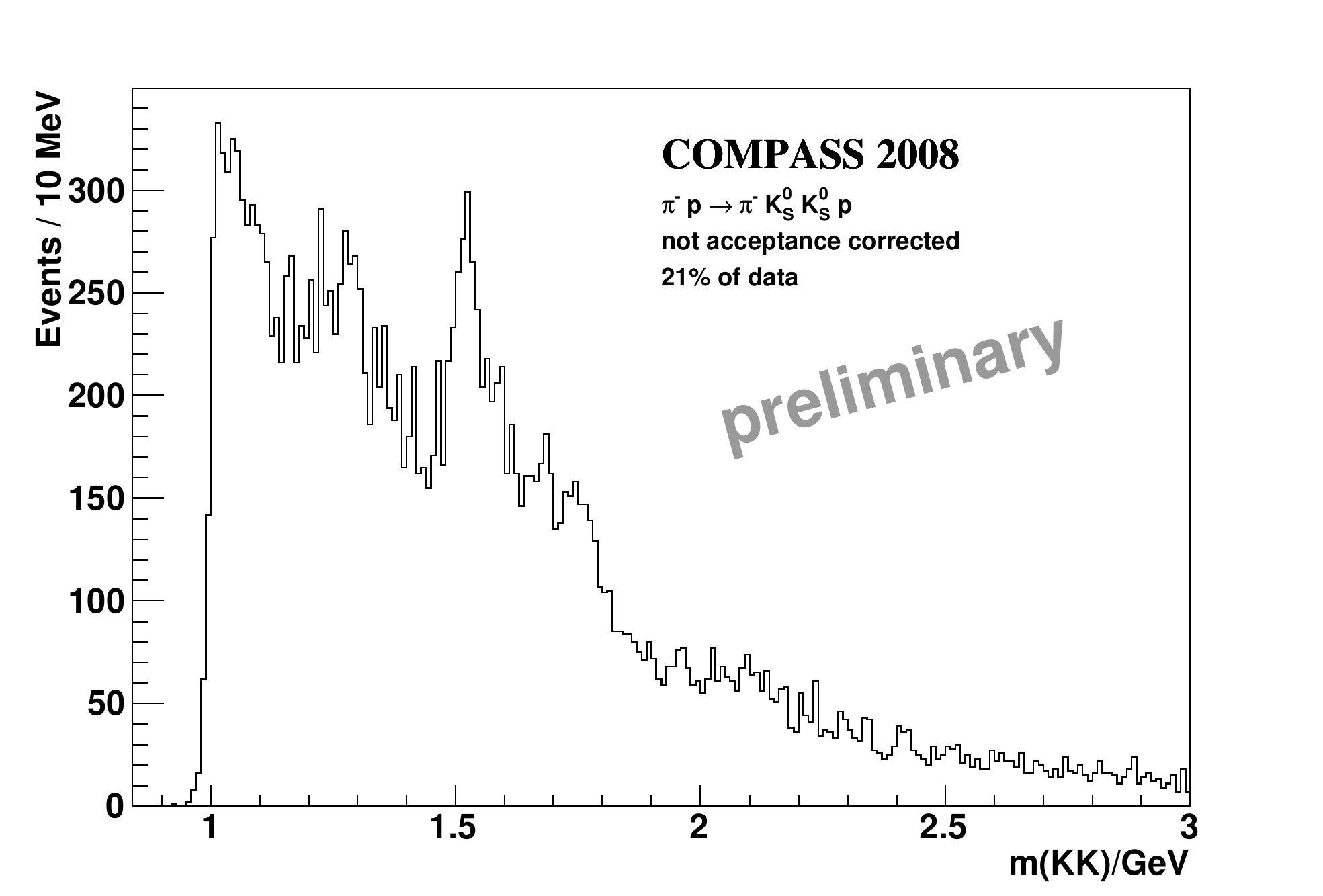}
    \caption{\em Invariant mass distribution of (Top) $\pi^-K^0_S$ and (Bottom) $K^0_SK^0_S$ sub-systems}
    \label{fig:Kneutral}
  \end{minipage}
\end{figure}

Figure~\ref{fig:Kcharged} and \ref{fig:Kneutral} show the invariant masses of the sub-systems of the studied $\pi K\bar K$ final states. While charged kaons are identified by the RICH detector, the neutral ones are reconstructed via their well resolved decay vertices. Both modes have a remarkable resemblance, pointing not only to $K*$(892) and higher mesons with strangeness but showing also structures around the glueball candidate $f_0$(1500).

\begin{figure}[h]
  \begin{minipage}[]{.48\textwidth}
    \centering
    \includegraphics[width=\textwidth]{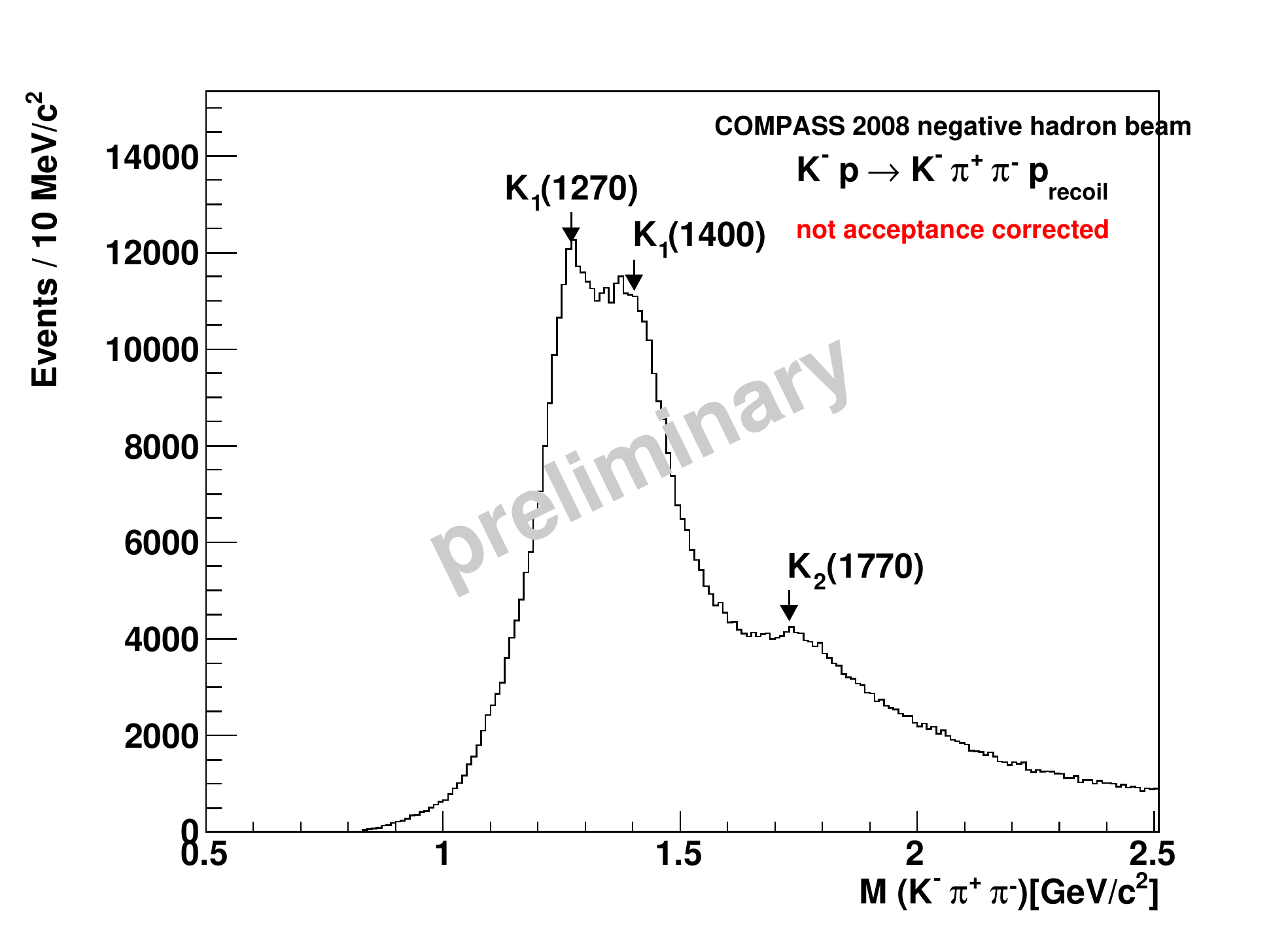}
    \caption{\em Invariant mass distribution of $K^-\pi^+\pi^-$ system}
    \label{fig:Kpippim}
  \end{minipage}
  \hfill
  \begin{minipage}[]{.48\textwidth}
    \centering
    \includegraphics[width=\textwidth]{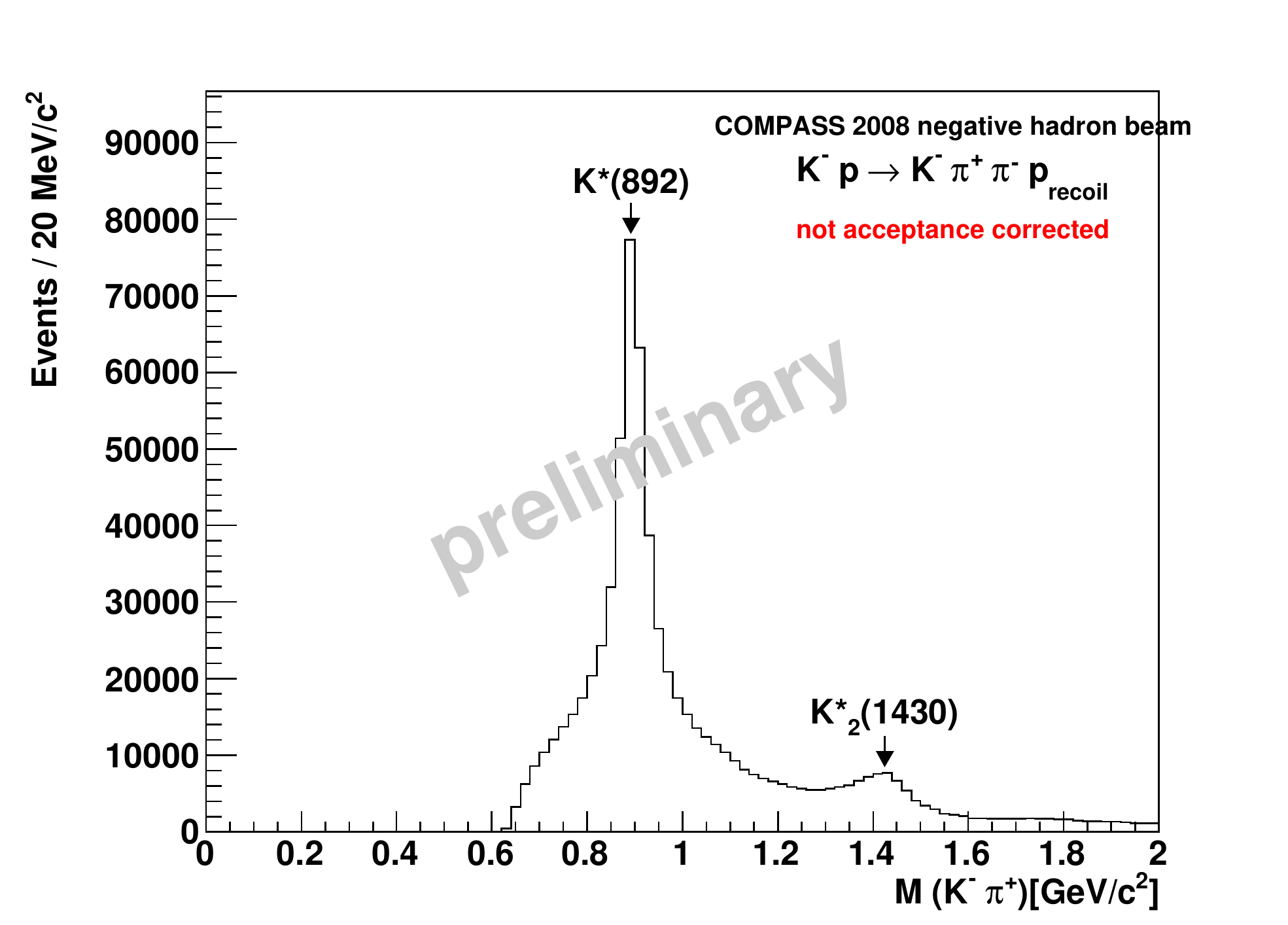}
    \caption{\em Invariant mass distribution of $K^-\pi^+$ sub-system}
    \label{fig:Kpip}
  \end{minipage}
\end{figure}

Further ongoing analyses use the kaon content of the negative beam to study kaon diffraction into $K^-\pi^+\pi^-$ final states, for example. Figure~\ref{fig:Kpippim} shows the total mass spectrum, where the $K_1$ double-structure is apparent. Including the isobars visible in Figure~\ref{fig:Kpip}, a dedicated partial-wave analysis for this channel is under preparation. Many states in this sector were previously observed by only one experiment\cite{wa03}, thus COMPASS can confirm their existence, exceeding the data set by at least a factor of two.

\newpage

\section{Baryon Spectroscopy}

Since the trigger system introduced no bias on the kinematics of the forward-going particles, the data recorded with positive hadron beam give a unique possibility to study diffractive dissociation of the beam protons. The protons in the liquid hydrogen target are assumed to be inert under the reaction. Exclusive events with one proton and a pair of either oppositely charged pions or kaons in the final state have been selected and will be the starting point for a dedicated partial-wave analysis. Hadron-induced reactions are complementary to the existing data from photo- and electro-production experiments like CBELSA or CLAS and may help to obtain a more complete picture of the baryon spectrum~\cite{kle10}. In particular poorly known parameters like widths and branching ratios of high-mass and high-angular-momentum states may become accessible.

\begin{figure}[h]
  \begin{minipage}[]{.48\textwidth}
    \centering
    \includegraphics[width=.95\textwidth]{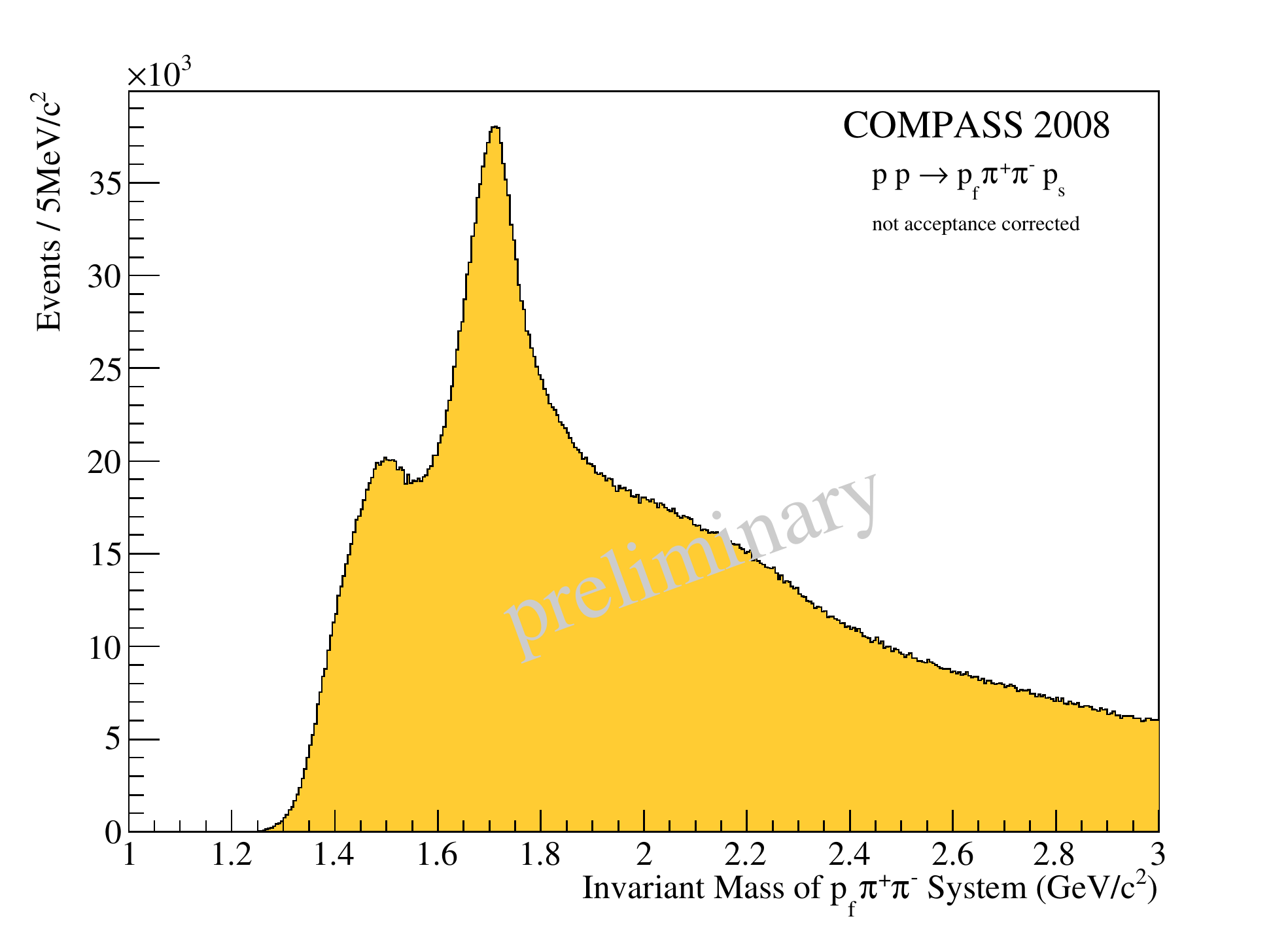}
    \caption{\em Invariant mass distribution of $p_f\pi^+\pi^-$ system}
    \label{fig:pfpippim}
  \end{minipage}
  \hfill
  \begin{minipage}[]{.48\textwidth}
    \centering
    \includegraphics[width=.95\textwidth]{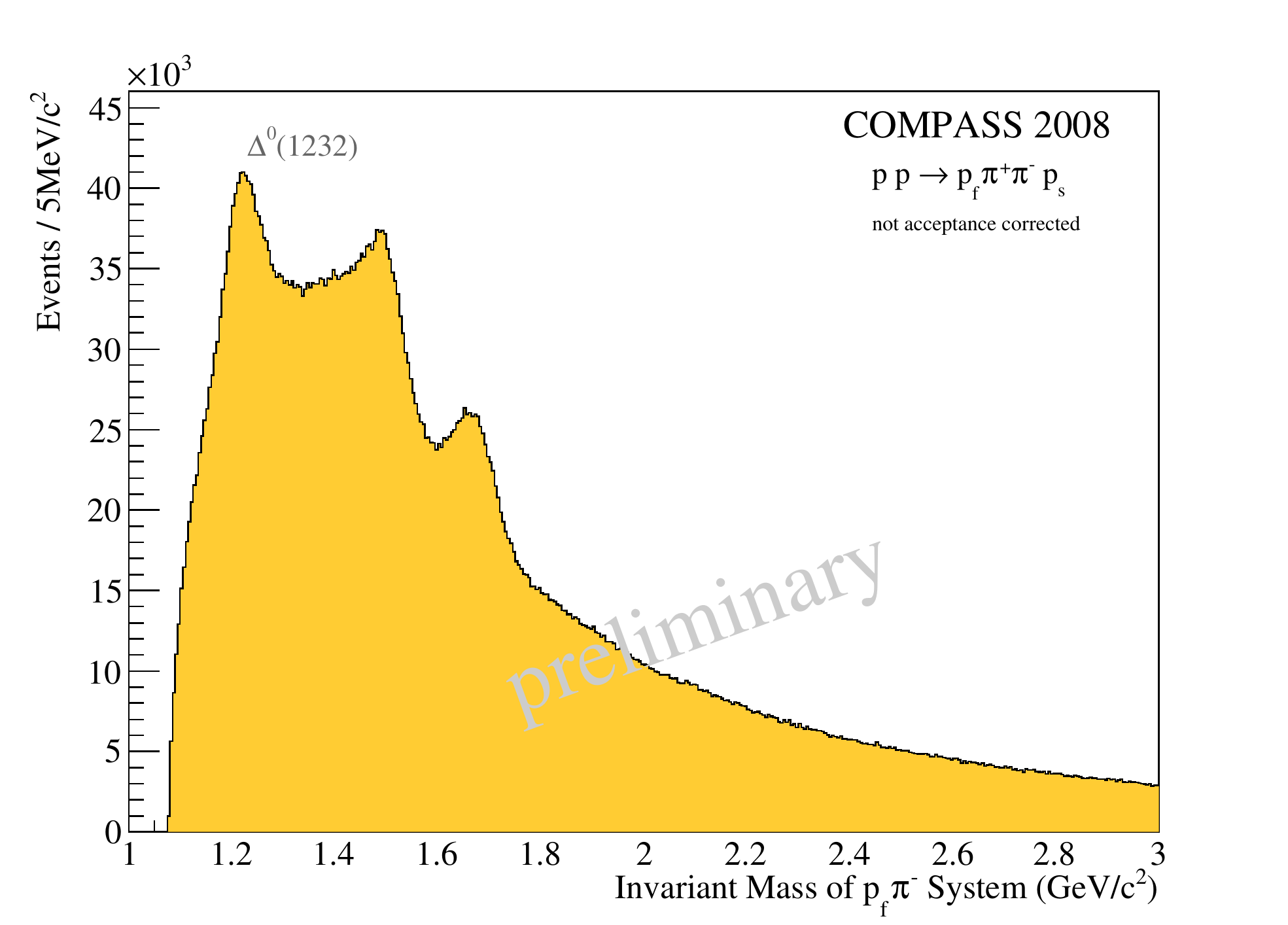}
    \caption{\em Invariant mass distribution of $p_f\pi^-$ sub-system}
    \label{fig:pfpim}
  \end{minipage}
\end{figure}

This excited proton spectrum visible in the invariant mass distribution of the $p_f\pi^+\pi^-$ system (cf.~Figure~\ref{fig:pfpippim}) is foreseen to be studied in detail by the means of partial-wave analysis. Few distinct structures can be observed at positions where there are several known $N^*$ and $\Delta$ resonances with $N\pi\pi$ decay modes.

The invariant mass spectrum of the $p_f\pi^-$ subsystem, depicted in Figure~\ref{fig:pfpim}, exhibits a distinct excited baryon spectrum, which actually constitutes part of the background for central production reactions. It features a prominent $\Delta^0$(1232)$P_{33}$ together with additional structures that are probably related to the $N$(1440)$P_{11}$, $N$(1650)$S_{11}$ and $\Delta$(1700)$D_{33}$. However, also here assignments based on the mass alone are ambiguous.

\begin{figure}[]
  \begin{minipage}[]{.48\textwidth}
    \centering
    \includegraphics[width=.95\textwidth]{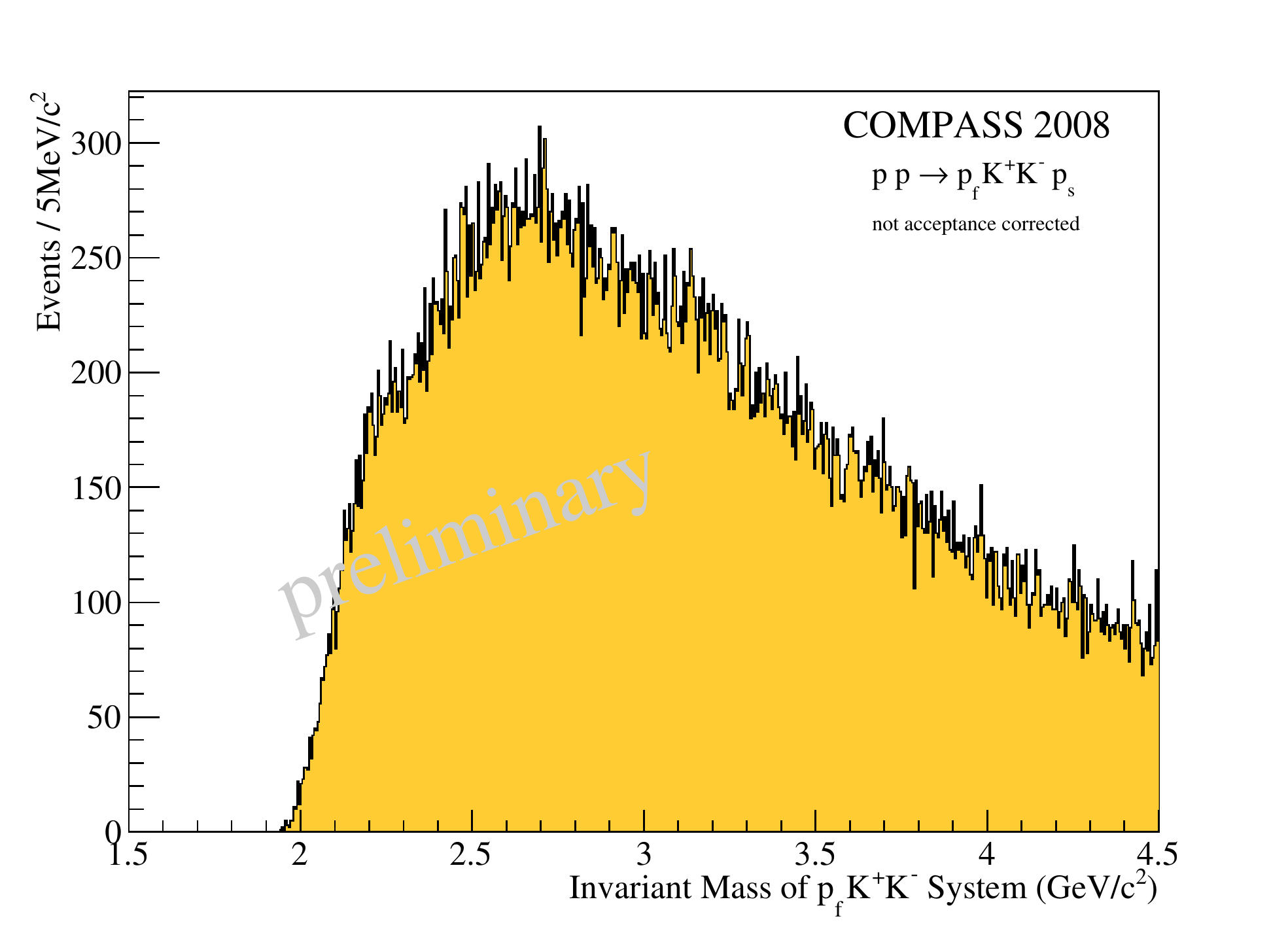}
    \caption{\em Invariant mass distribution of $p_fK^+K^-$ system}
    \label{fig:pfKpKm}
  \end{minipage}
  \hfill
  \begin{minipage}[]{.48\textwidth}
    \centering
    \includegraphics[width=.95\textwidth]{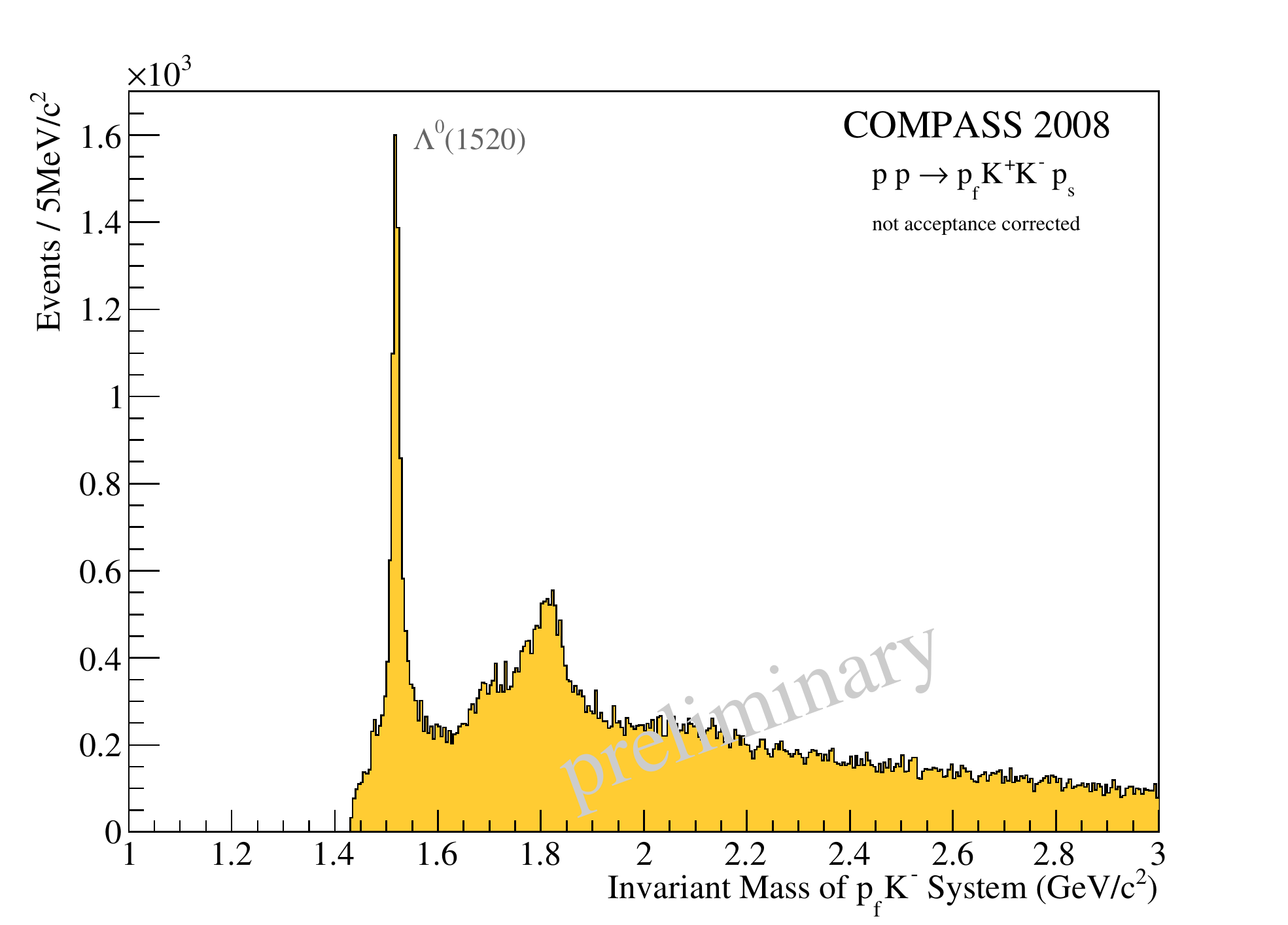}
    \caption{\em Invariant mass distribution of $p_fK^-$ sub-system}
    \label{fig:pfKm}
  \end{minipage}
\end{figure}

A different aspect of the baryon spectrum becomes accessible when the pions are replaced by kaons in the event selection described above. However, the number of events is considerably lower and therefore the identification of resonances is more difficult. While no special features can be seen in the three-particle invariant mass spectrum (cf.~Figure~\ref{fig:pfKpKm}), the subsystems do show interesting structures. A sharp baryon resonance, the $\Lambda$(1520)$D_{03}$, can be found in the invariant mass spectrum of the $p\,K^-$ combination (cf.~Figure~\ref{fig:pfKm}). Higher baryon excitations with strangeness are visible for example around $1.7$ and $1.8\,\mathrm{GeV}/c$, although less pronounced. 

\section{Conclusion and Outlook}

After the observation of a spin exotic $J^{PC}=1^{-+}$ state consistent with $\pi_1$(1600) in the data recorded for the 2004 pilot run, COMPASS collected a world leading data set of negatively and positively charged hadron beams ($p$,$\pi^\pm$, $K^\pm$) on both nuclear and liquid hydrogen targets. This will give access to various open questions in light-meson and baryon spectroscopy to masses up to $3\,\mathrm{GeV}/c^2$ with unprecedented accuracy. The partial-wave analysis tools were elaborated and tested on real data and Monte Carlo simulation, the latter necessary to understand the spectrometer acceptance in detail. Several different analyses are currently underway, out of which only few could be highlighted here. COMPASS has excellent prospects to shed more light into the field of exotic mesons within the next years.

\newpage

\section*{Acknowledgements}

This work is supported by the German Bundesministerium f\"ur Bildung und Forschung, the Maier-Leibnitz-Labor der LMU und TU M\"unchen, and the DFG Cluster of Excellence \textit{Origin and Structure of the Universe}.

\end{document}